\documentclass[prd,preprintnumbers,twocolumn,floatfix,footinbib,superscriptaddress]{revtex4}
\usepackage{amsmath,amssymb}
\usepackage{graphicx}

\newcommand{\tmdpdf}{\ensuremath{\tilde{F}}}
\newcommand{\trans}[1]{\ensuremath{{\bf #1}_T}}

\newcommand{\gammae}{\ensuremath{\gamma_{\rm E}}}
\newcommand{\subT}[1]{\ensuremath{#1_{\rm T}}}

\def\slash#1{\ooalign{$\hfil/\hfil$\crcr$#1$}}

\begin{document}
\title{TMD Parton Distribution and Fragmentation Functions with QCD Evolution}

\date{\today}

\preprint{NIKHEF-2011-001}

\author{S.~Mert Aybat}
\email{maybat@nikhef.nl}
\affiliation{Nikhef Theory Group, \\ Science Park 105, 1098XG Amsterdam, The Netherlands}
\affiliation{Department of Physics and Astronomy,\\ Vrije Universiteit Amsterdam,\\ NL-1081 HV Amsterdam, The Netherlands}
\author{Ted C.~Rogers}
\email{trogers@few.vu.nl}
\affiliation{Department of Physics and Astronomy,\\ Vrije Universiteit Amsterdam,\\ NL-1081 HV Amsterdam, The Netherlands}
\date{\today}

\begin{abstract}
We assess the current phenomenological status of  
transverse momentum dependent (TMD) parton distribution functions (PDFs) and fragmentation 
functions (FFs)
and study the effect of consistently including perturbative QCD (pQCD) evolution. 
Our goal is to initiate the process of establishing reliable, QCD-evolved parametrizations for the TMD PDFs and TMD FFs that
can be used both to test TMD-factorization and to search for evidence of the breakdown of TMD-factorization that is expected 
for certain processes.
In this article, we focus on spin-independent processes because they provide the simplest illustration 
of the basic steps and can already be used in 
direct tests of TMD-factorization. 
Our calculations are based on the Collins-Soper-Sterman (CSS) formalism, supplemented by 
recent theoretical developments which have clarified the precise definitions of the TMD PDFs and TMD FFs 
needed for a valid TMD-factorization theorem.  
Starting with these definitions, we numerically generate evolved TMD PDFs and TMD FFs using as input 
existing parametrizations for the collinear PDFs, collinear FFs, non-perturbative factors 
in the CSS factorization formalism, and recent fixed-scale fits.
We confirm that evolution has important consequences, both qualitatively and quantitatively, and argue that it
should be included in future phenomenological studies of TMD functions.
Our analysis is also suggestive of extensions to  
processes that involve spin-dependent functions such as the Boer-Mulders, Sivers, or 
Collins functions, 
which we intend to pursue in future publications. 
At our website~\cite{webpage} we 
have made available the tables and 
calculations 
needed to obtain the 
TMD 
parametrizations presented herein.
\end{abstract}
%\keywords{}
%\pacs{}
\maketitle

\section{Introduction}
\label{intro}
The factorization theorems of pQCD have been instrumental 
in the successful application of QCD theory to phenomenology.
The standard 
collinear factorization formalism~\cite{Collins:1989gx} makes use of ``integrated''
PDFs and FFs which depend only on 
a single longitudinal momentum fraction, while the small momentum components, 
including the \emph{transverse} components, are integrated over in the definitions.
The integrated PDFs and FFs have 
consistent operator definitions in QCD, with appealing interpretations in terms of parton model concepts.
However, the standard collinear factorization formalism relies on approximations
that are only valid for sufficiently inclusive observables.  
In order to address many of 
the issues now at the forefront of research in
QCD and its role in hadron structure, pQCD factorization 
must be extended to situations where the 
usual 
approximations are not appropriate.

A 
transverse momentum dependent
\emph{TMD-factorization} formalism goes beyond the standard factorization framework 
by allowing the PDFs and FFs to depend on intrinsic transverse momentum in addition to the usual momentum 
fraction variables.  
As such, different sets of approximations are needed in the factorization proofs.
The PDFs and FFs in a TMD-factorization formalism 
are referred to as TMD PDFs and TMD FFs (they are also 
called ``unintegrated'' or `` $k_T$-dependent'') to distinguish 
them from the more familiar integrated correlation functions of collinear factorization. 
Henceforth, we will refer to TMD PDFs and TMD FFs collectively as ``TMDs''.

The role of the intrinsic transverse momentum carried by partons 
in high energy collisions is becoming increasingly central in discussions of how to probe the details of hadronic
structure in high energy collisions.  
The usefulness of the TMD concept is evident from the large variety of situations where it makes an appearance.  
Generally, TMD-factorization is needed to describe processes that are sensitive to intrinsic parton transverse momentum.
The Drell-Yan (DY) process, 
single-inclusive deep inelastic scattering (SIDIS), and back-to-back hadron production in 
electron-positron annihilation at small transverse momentum are all classic examples of where TMD-factorization 
formulae are frequently used.
More recently, TMD PDFs and FFs have been under intense study as objects that carry information about the spin structure 
of hadrons; the Boer-Mulders, Sivers, and 
pretzelosity
functions are all specific examples of TMD PDFs, while the Collins function 
is an example of a TMD FF.
For a recent review of TMDs in spin physics, see Ref.~\cite{Barone:2010zz}.
In very high energy (small-$x$) resummation physics, where there is a lack of $k_T$-ordering, the TMD gluon distribution is especially important, and 
similar issues must be dealt with.
Finally, TMD functions are useful tools in the construction of Monte Carlo event generators, where
the details of final state kinematics are of interest. 

While 
TMDs can potentially provide a much 
deeper understanding of QCD and hadron structure, the theoretical 
framework of TMD-factorization is much more complicated than 
the more standard collinear factorization.
In derivations of collinear factorization, there are important cancellations that occur 
\emph{after} integrations of parton momentum are carried out.
With TMD-factorization, the integrals over parton transverse momentum are left undone in 
the definitions of the TMDs, 
and contributions that would ordinarily cancel in a collinear factorization treatment must be accounted for.
Collins, Soper and Sterman (CSS) constructed a TMD-factorization formalism~\cite{CSS1,CSS2,CSS3} that deals with the
main complications of transverse momentum dependence, 
and provides a systematic treatment of pQCD over the full range of transverse momentum. 
The CSS formalism has proven highly successful in 
specific phenomenological applications
such as in the calculation of the 
transverse momentum distributions in DY processes (see, for example,~\cite{Davies:1984sp,Balazs:1997xd}), and is also 
well-suited for the production of back-to-back particles in $e^{+}e^{-}$ annihilation.
The same methods are needed for the discovery of new particles like the standard model Higgs boson~\cite{higgs,higgs2,higgs3}.
Furthermore, extensions of the CSS TMD-factorization formalism have been derived 
for SIDIS~\cite{Meng:1995yn,Nadolsky:1999kb,Ji:2004wu}, and including spin in Ref.~\cite{Ji:2004xq}.

However, the most common methods for 
applying the CSS formalism are not ideally suited for 
studies that are specifically oriented toward  
understanding the TMD PDFs and TMD FFs themselves.
Furthermore, the relationship between the full pQCD treatment of factorization and 
parton-model intuition has remained much less clear in TMD-factorization than in collinear factorization. 
This has led to considerable confusion about how the study of TMDs should be approached in pQCD.
That confusion is especially apparent from a comparison between current applications of TMD-factorization and collinear factorization: 
While there have been extensive programs dedicated 
to parametrizing and evolving 
the integrated PDFs and FFs 
(making them indispensable and portable tools for 
phenomenology), a generally agreed upon framework for 
dealing with TMDs in an analogous way has not yet been established.  
More disturbingly, there has been a persistent lack of 
clarity or agreement regarding the definitions of the TMDs.
A suitable set of definitions must be dictated by the requirements of 
factorization and universality, but the most 
common and
naive definitions lead to 
inconsistencies, including unphysical divergences.   

Over roughly the past decade, there has been steady progress toward 
a better understanding of what is needed ~\cite{Collins:1999dz,Collins:2000gd,Henneman:2001ev, 
Belitsky:2002sm,Boer:2003cm,C03,Hautmann:2007cx,Collins:2007ph,Cherednikov:2010uy,Collins:2008ht}.
The issue of finding the right definitions 
has now been especially brought into focus by the recent work of Collins~\cite{collins,collins_trento}. 
The definitions for the TMDs in Ref.~\cite{collins} are uniquely determined
by the requirements of factorization, maximal universality, and internal 
consistency. 
(By ``maximal universality'' we mean that the same correlation functions appear 
in a large number of processes.)
The confusion over definitions therefore appears to be solved.
With the new definitions, the implementation of evolution via the 
CSS formalism is not modified significantly from earlier treatments which
means that existing parametrizations of the non-perturbative parts can still be used.
However, they can now be understood as contributions to separate, QCD-evolved TMDs.  
Moreover, as we will discuss in the next two sections, the 
new definitions have a much more direct relationship with more 
intuitive, parton model based ways of viewing TMD-factorization.

There has been much recent work devoted to 
parametrizing TMDs by assuming a 
parton model picture of TMD-factorization and directly fitting
cross section calculations to experimental 
data~\cite{Anselmino:2004nk,Anselmino:2005sh,Anselmino:2007fs,Anselmino:2008sga,Anselmino:2009st,Schweitzer:2010tt}.
This approach to TMD phenomenology is often called the \emph{generalized parton model} (GPM)~\cite{D'Alesio:2007jt}.

In addition, by working with non-perturbative models, it is possible to study the general properties of the 
TMDs and their relationships with each other.  (See Ref.~\cite{Avakian:2009jt} and references therein for an overview 
of this subject.)  A famous example is the illustration via model calculation that the Sivers function is non-vanishing 
in SIDIS~\cite{Brodsky:2002cx,Brodsky:2002rv}.

However, most efforts to parametrize or model TMDs, particularly the spin-dependent TMDs, have ignored the role 
of evolution. For the unpolarized TMD PDFs there have been detailed implementations 
of evolution (e.g.~\cite{Ladinsky:1993zn,Landry:2002ix}) 
However, even in the unpolarized case, the identification of the separate evolved TMDs and their 
relationship with the fundamental definitions in a complete treatment of factorization has remained unclear.
The main purpose of the present article is to initiate the process of consistently 
including QCD evolution in parametrizations and models
of TMDs by following the definitions in Ref.~\cite{collins}. 
Addressing evolution is now an especially pressing task, given the very 
wide range of energy scales set to be probed in experiments in the near future, from 
Jefferson Lab (JLab) 
to the LHC.
It has already been shown in Refs.~\cite{Boer:2001he,Boer:2008fr} that evolution (in the form of Sudakov suppression)
should be expected to be large. 
In our analysis, we will illustrate how this follows from the evolution of the individual TMDs.
We also aim to facilitate the future implementation of evolution in studies of TMDs by clarifying the 
relationship between the parton model description of TMD-factorization and the CSS formalism.
To this end, we have made computations available at~\cite{webpage} that illustrate how to 
obtain evolved TMDs given 
a choice of  
non-perturbative starting input. 

An alternative approach to probing spin effect is to consider higher twist collinear 
functions such as the Qiu-Sterman function~\cite{Qiu:1991wg}.  By taking transverse momentum moments of 
cross sections it is possible to relate TMDs to these higher twist functions~\cite{Boer:2003cm}.
See Ref.~\cite{Kang:2010xv} for recent work on the evolution of weighted spin-dependent correlation functions.

Establishing reliable, evolved TMDs is also important for testing factorization and 
searching for instances of factorization breaking.
Discussions of non-universality or factorization breaking often bring in the concept of 
Wilson lines (or gauge-links), and the 
process dependent properties that can be associated with them.
Already when comparing the Sivers function in SIDIS and the DY process, 
one must account for a well-known sign flip that is due to the reversed direction of the Wilson lines in 
these two processes~\cite{Collins:2002kn}.
More recently, it has been shown that obtaining consistent Wilson lines is 
even more problematic in the hadro-production of hadrons or jets ($H_1 + H_2 \to H_3 + H_4 + X$).
In such cases that require TMD-factorization, it was found that \emph{at a minimum} the Wilson lines
for separate TMDs are highly complex and process dependent~\cite{factbreaking1,factbreaking2,factbreaking3}.  
It was later shown in Ref.~\cite{factbreaking4}  
that TMD-factorization generally breaks down completely in 
the hadro-production of hadrons.  That is, separate
TMDs cannot even be defined for each external hadron 
regardless of what Wilson lines are used in the definitions. 

The complication in the case of the complete breakdown of factorization 
is caused by a failure of the usual 
gauge invariance/Ward identity arguments that are needed in a factorization proof.
A confirmation of this effect 
would 
point to interesting new features of strong interaction physics, given that the 
breakdown of factorization 
is
in just the range of kinematics where factorization 
would naively be expected to hold.  
Calculations in a GPM framework~\cite{Boer:2007nd,Boer:2009nc,D'Alesio:2010am} will be 
needed 
for making predictions that can be compared with experiment to test factorization and/or 
search for factorization breaking.  
Furthermore, computations using the methods
in, for example, Refs.~\cite{Gelis:2008rw,Balitsky:2001gj,Dominguez:2010xd} can potentially help to quantify and better understand the 
factorization breaking mechanism.  
It may soon be possible to find experimental evidence for factorization
breaking, particularly in the analysis
of RHIC data.  (See, for example, the recent analysis of Ref.~\cite{Abelev:2007ii,Adare:2009js,Adare:2010bd,Adare:2010yw}.) 
However, definitive tests of factorization or factorization breaking can only become possible
with a more precise determination of the TMDs.
Another 
motivation for this article is therefore to 
begin the determination of TMD parametrizations that 
should be used in tests of factorization in the case of spin-independent hadro-production
of back-to-back hadrons.
Finally, in this paper we include some details of the calculation of 
the evolution of the TMD PDFs that did not appear in Ref.~\cite{collins}

The paper is organized as follows: in Sect.~\ref{sec:overview}, a brief 
background of TMD-factorization is provided,
and the main complications that arise in the context of pQCD are discussed. In 
Sect.~\ref{sec:setupandnotation}, we set up the notation,  
and in Sect.~\ref{sec:defins} we explain the definitions 
of TMD correlation functions. In Sect.~\ref{sec:evolved}, we 
discuss the evolution of the TMDs in terms of the CSS formalism.
We apply evolution to existing unpolarized quark TMD fits  
in Sect.~\ref{sec:impevol}, and we present some numerical results.   
We conclude with an overview and a discussion of future work in Sect.~\ref{sec:conclusion}.
In the appendices, we provide some details of the perturbative parts of our calculations.

\section{The Different Pictures of TMD-Factorization}
\label{sec:overview}

In this section we expand on the general remarks in the 
introduction by providing a very schematic 
overview of the different ways TMD-factorization is approached 
in phenomenological applications.
We discuss the relationship between parton 
model intuition and full pQCD while emphasizing the complications that can arise.  
We explain the potential for confusion when evolution and soft factors are  
included, and how this is solved by 
the use of 
fully consistent definitions for the TMDs.

\subsection{The Generalized Parton Model}
\label{sec:gpm}

We start by considering the simplest parton model picture 
of high energy collisions. 
There, the 
concept of a TMD-factorization formula becomes very intuitive and 
easy to state.  
The cross section is simply 
a partonic sub-process, 
folded with TMD PDFs and TMD FFs.  In SIDIS, for example, the 
hadronic tensor is written as
\begin{multline}
\label{eq:parton1}
W^{\mu \nu} =
\sum_f \left| \mathcal{H}_f(Q)^2 \right|^{\mu \nu} 
\, \times \\ \int \, d^2 {\bf k}_{1T} \, d^2 {\bf k}_{2T} \, F_{f/p}(x,{\bf k}_{1T}) \, D_{h/f}(z,z {\bf k}_{2T}) 
\times \\ \times \delta^{(2)}({\bf k}_{1T} + {\bf q}_T - {\bf k}_{2T}).
\end{multline} 
Here $| \mathcal{H}_f(Q)^2 |^{\mu \nu}$ 
describes the hard partonic sub-process, $\gamma^{\ast} q \to q$, 
for scattering off a quark of flavor $f$ as a function of the hard scale $Q$. 
(It also includes any overall factors needed to make the left side a proper hadronic tensor.) 
The size of ${\bf q}_T$ is a measure of the non-collinearity in the process.
Within the parton model, the TMDs 
$F_{f/p}(x,{\bf k}_{1T})$ and $D_{h/f}(z,z {\bf k}_{2T})$ have simple 
probabilistic interpretations;
$F_{f/p}(x,{\bf k}_{1T})$, for example, is the probability density for finding a quark of flavor $f$ with momentum fraction 
$x$ and transverse momentum ${\bf k}_{1T}$ inside the proton.
Equation~(\ref{eq:parton1}) is closely analogous to the standard \emph{collinear}
factorization theorem of inclusive processes~\cite{Collins:1985ue}.  
The only difference is that the TMD PDFs and FFs are allowed to carry intrinsic transverse momentum.

The intuitive approach to TMD-factorization embodied by Eq.~(\ref{eq:parton1})
forms the basis of much current TMD phenomenology.
However, derivations of TMD-factorization 
order-by-order in real pQCD involve complications that are 
not immediately apparent in parton model reasoning.
Indeed, some aspects of partonic intuition can quickly lead to incorrect results if taken too literally. 
Furthermore, without a complete derivation of factorization, the issue of 
universality (or non-universality) of the TMDs in Eq.~(\ref{eq:parton1}) is much less clear.

\subsection{Divergences and Soft Factors}
\label{sec:divsoft}

One issue is the appearance of extra divergences. 
In addition to the standard UV divergences associated with the 
renormalization of the theory, the TMD correlation functions also contain so-called ``light-cone''
divergences.
They arise if the Wilson lines (gauge links) 
in the definitions of the TMDs point in exactly light-like directions.
We should note that the same light-cone singularities are not present 
in the ordinary collinear correlation functions because they cancel in 
a sum over final state interactions, which is possible due to integrations 
over the loop momenta, including the $k_T$ integrals.
Light-cone divergences   
correspond to gluons moving with infinite rapidity in the direction \emph{opposite}
the containing hadron, and are not regulated by the use of an infrared cutoff, so they 
amount to a real inconsistency in the definitions of TMDs.
The most naive definitions, therefore, are invalid, and modifications are needed for a 
reliable factorization theorem.
(See Refs.~\cite{C03,Collins:2008ht} and references therein 
for a review of this and other subtleties involved in defining consistent PDFs.)
The light-cone divergences need to be regulated, typically by tilting the Wilson lines 
slightly away from the exactly light-like directions.  

Furthermore, in a TMD-factorization formula the role of soft gluons 
becomes important.  (In this paper, ``soft'' refers to nearly on-shell gluons 
with rapidity intermediate between the rapidities of 
the colliding hadrons and produced jets.) 
They imply that another correlation function, a separate soft 
factor, should be inserted into factorization formulae like Eq.~(\ref{eq:parton1}), in addition to the usual TMD PDFs and FFs.
Already, the appearance of a soft factor seems to contradict the basic parton-model intuition.

The complications listed above, as well as a consistent matching to collinear 
factorization at large transverse momentum, are accounted for in the CSS formalism~\cite{CSS1,CSS2,CSS3}.
With the Wilson lines tilted to remove 
light-cone 
divergences, the factorization formula 
acquires new arbitrary parameters.  
Predictive power is then recovered 
by a kind of generalization of renormalization group techniques.  The resulting evolution equations may be thought of as
describing the variation of the TMDs and the soft factors with changes in the degree of tilt of the Wilson lines. 
Physically, this corresponds roughly to a variation with respect to a cutoff on the phase space available for recoil against soft gluons.

\subsection{Confusion Over TMD Definitions}
\label{sec:confusing}

While the CSS formalism has been very useful for past phenomenological studies, the usual implementations   
bear little surface resemblance to the generalized partonic picture we started out with in Eq.~(\ref{eq:parton1}).
For example, in Ref.~\cite{Landry:2002ix} (and similar applications of the CSS formalism to the DY 
process) the effects of evolution are gathered into separate factors, and it is not clear how they 
relate to separate TMD PDFs.  In other treatments (e.g.~\cite{Ji:2004wu,Bacchetta:2006tn}), 
factorization formulae for SIDIS
are provided which contain explicit evolved TMDs, but they
also involve separate explicit soft factors, and the hard part has explicit dependence on 
light-cone divergence cutoff parameters.  
Moreover, given the general observations 
that are reviewed in
Refs.~\cite{C03,Collins:2008ht,Stefanis:2010vu}, it is questionable 
whether the most commonly quoted definitions of the TMDs are even  
fully consistent.

The original work of Collins and Soper~\cite{CSS1} used a non-light-like axial gauge 
to regulate the rapidity divergences.  Later, Collins and Hautmann~\cite{Collins:1999dz,Collins:2000gd,C03} proposed definitions in which the 
main legs of the Wilson lines are light-like, but 
in which there is a division by a special type of soft factor
which cancels the rapidity divergences.  However, these definitions continue to suffer from problems,  
including the appearance of badly divergent Wilson line self-energy contributions 
as discussed recently in Ref.~\cite{Collins:2008ht} and 
also utilized in the treatment of TMD PDFs in Ref.~\cite{Cherednikov:2010uy}. 
While many of these issues have typically been discussed in the context of TMD PDFs, the same 
problems arise in the treatment of FFs.

Finally, it has remained unclear how the TMDs that have been used in 
past applications of evolution and the CSS formalism 
are related to the TMDs of other approaches, such as those 
based more on generalized parton model pictures. 
In parametrizations of TMDs, the role of the soft factor is 
often not explicitly included and evolution is ignored.
Many other theoretical TMD studies continue to quote definitions with exactly light-like 
Wilson lines. 
Knowledge of the operator definitions for the TMDs is also 
needed for lattice TMD calculations~\cite{Musch:2009ku,Musch:2010ka}, and 
in model calculations.    
Clearly, a more unified treatment of TMD-factorization is necessary in order 
to bring together 
these different approaches to the study of TMDs.

\subsection{Consistent Definitions, TMD-factorization and Evolution}
\label{sec:TMDandevol}

What is needed, in addition to fully consistent TMD definitions, is a formulation that retains as much as possible the basic 
factorized structure of Eq.~(\ref{eq:parton1}), but which appropriately includes evolution and the effects of soft factors.
Ideally, the situation should be 
closely 
analogous to what already exists for collinear factorization.
Namely, it should be possible to clearly identify consistent and universal TMDs that can be tabulated or parametrized
and then reused in formulae like Eq.~(\ref{eq:parton1}) to make predictions for a wide variety of processes.
Fortunately, this has 
been achieved 
in the recent work of Ref.~\cite{collins}.  
There, the factorization formula for SIDIS takes the form:
\begin{multline}
\label{eq:parton3}
W^{\mu \nu} =
\sum_f | \mathcal{H}_f(Q;\mu)^2 |^{\mu \nu} \, \times \\ \int \, d^2 {\bf k}_{1T} \, d^2 {\bf k}_{2T} \, 
F_{f/p}(x,{\bf k}_{1T};\mu;\zeta_F) \, D_{h/f}(z,z {\bf k}_{2T};\mu;\zeta_D) 
\times \\ \times \delta^{(2)}({\bf k}_{1T} + {\bf q}_T - {\bf k}_{2T}) \\
+ Y(Q,\trans{q}) + \mathcal{O}((\Lambda / Q)^a).
\end{multline}  
The first term on the right-hand side of this equation 
(responsible for the low-${\bf q}_T$ behavior) has exactly the structure of the partonic TMD-factorization 
formula in Eq.~(\ref{eq:parton1}), apart from the scale dependence denoted by $\mu$, $\zeta_F$ and $\zeta_D$.
The arguments $\zeta_F$ and $\zeta_D$ will be discussed more in the 
explanation of the TMD definitions in Sect.~\ref{sec:defins}.  They are left over from the need to regulate 
light-cone divergences, and should obey $\sqrt{\zeta_F \zeta_D} \sim \mathcal{O}(Q^2)$. 
In terms of more familiar variables, they are defined as:
\begin{equation}
\label{eq:zetaF}
\zeta_F = 2 M_p^2 x^2 e^{2(y_P - y_s)}
\end{equation}
and 
\begin{equation}
\label{eq:zetaD}
\zeta_D = 2 (M_H^2 /  z^2) e^{2(y_s - y_h)}.
\end{equation}
Here, $x$ and $z$ are the usual Bjorken scaling and fragmentation variables, $M_p$ is the proton mass and $M_h$ is the 
mass of the produced hadron.  The rapidities of the proton and produced hadron are $y_p$ and $y_h$ respectively.  The 
rapidity $y_s$ is an arbitrary low-rapidity cutoff parameter that separates partons with large forward rapidity (in the 
proton direction) from backward rapidity (in the produced hadron direction).  Variations of these functions with $y_s$ will be determined by 
the evolution equations.

The scale $\mu$ is the standard renormalization group (RG) scale.
The TMD correlation functions, $F_{f/p}(x,{\bf k}_{1T};\mu;\zeta_F)$ and $D_{h/f}(z,z {\bf k}_{2T};\mu;\zeta_D)$, have 
definite and consistent operator definitions.  
They include the effects from soft gluons in such a way that no soft factor appears 
explicitly in Eq.~(\ref{eq:parton3}). 
Evolution can be implemented on $F_{f/p}(x,{\bf k}_{1T};\mu;\zeta_F)$ 
and $D_{h/f}(z,z {\bf k}_{2T};\mu;\zeta_D)$ independently, 
and the basic steps closely follow the usual 
CSS approach.
We will discuss the definitions more in the next section, but for now we mention that they 
solve most of the theoretical problems summarized in Refs.~\cite{C03,Collins:2008ht} and Sect.~\ref{sec:confusing}, including the appearance
of light-cone divergences and Wilson line self-interactions. 

The term, $Y(Q,q_T)$, accounts for the large-${\bf q}_T$ dependence of the cross section, where 
the approximations needed for TMD-factorization break down.  
There, collinear factorization becomes the 
appropriate framework.  The error term is suppressed by $(\Lambda / Q)^a$ where $a > 0$.
The first term on the right side of Eq.~(\ref{eq:parton3}) is valid up to corrections of order $(q_T / Q)^a$, 
but the $Y(Q,\trans{q})$ is needed for a valid treatment of factorization over the full range of ${\bf q}_T$.

The derivation of Eq.~\eqref{eq:parton3} within pQCD factorization, with consistent 
definitions for the TMDs, is an important breakthrough because it connects TMD 
studies from a GPM framework with formal QCD and clarifies the meaning of TMD evolution.
We will use Eq.~(\ref{eq:parton3}), along with the associated 
definitions for the TMDs from Ref.~\cite{collins}, to obtain momentum space fits for use in phenomenology.
The non-perturbative input  
can be obtained from already existing models or fits made at
fixed scales.  For the TMD PDFs, much information about the non-perturbative input 
is already available from fits that use the standard ${\bf b}_T$-space formulation of the CSS formalism in 
the DY process.

\section{Setup and Notation}
\label{sec:setupandnotation}

We start by setting up the basic notation.  
In our convention 
for light-cone variables, a four-vector $V^\mu = (V^+,V^-,{\bf V}_T)$ has
components,
\begin{eqnarray}
V^\pm = \frac{V^0 \pm V^z}{\sqrt{2}} \nonumber\\
{\bf V}_T = (V^x,V^y).
\end{eqnarray}
The $z$-component picks out the forward direction.  Note that $V^2 = 2 V^+ V^- - {\bf V}_T^2$.

For the processes we are interested in, there 
are always two relevant light-like directions 
which we label $u_{\rm A}$ and $u_{\rm B}$ and define to be: 
\begin{equation}
\label{eq:LLdir}
u_{\rm A} = (1,0,{\bf 0}_t) \qquad u_{\rm B} = (0,1,{\bf 0}_t).
\end{equation} 
In the SIDIS example, $u_{\rm A}$ and $u_{\rm B}$ characterize 
the directions of the incoming proton and the produced jet.
A Wilson line from a coordinate $x$ to $\infty$ along the direction of a four-vector $n$ is defined as usual:
\begin{equation}
\label{eq:wildef}
W(\infty ,x;n) = P \exp \left[- ig_0 \int_0^\infty d s \; n \cdot A_0^a (x + s n) t^a \right].
\end{equation}
In these definitions, the bare fields and couplings are used, $P$ is a path-ordering operator,
and $t^a$ is the generator for the gauge group in the fundamental representation, with color index $a$.

As discussed in the previous section, light-cone divergences must
be regulated by tilting the direction of the Wilson line away from the exactly 
light-like 
direction.
Therefore, we need to define 
another set of vectors $n_{\rm A}$ and $n_{\rm B}$ 
analogous to Eq.~(\ref{eq:LLdir}) but slightly tilted, so that they 
have rapidities $y_A$ and $y_B$:
\begin{equation}
\label{eq:nLLdir}
n_{\rm A} = (1,-e^{-2 y_{\rm A}},{\bf 0}_t) \qquad n_{\rm B} = (-e^{2 y_{\rm B}},1,{\bf 0}_t).
\end{equation}
Note that the tilted Wilson line directions are space-like, $n_{\rm A}^2 = n_{\rm B}^2 < 0$.
The use of space-like directions for the Wilson lines ensures 
maximum universality for the definitions of the TMDs, as explained in Ref.~\cite{Collins:2004nx}. 
In all of our calculations,
$\mu$ is the standard $\overline{{\rm MS}}$ mass scale in dimensional regularization and the 
dimensional regularization parameter $\epsilon$ is defined in the standard way as $2 \epsilon = 4 - d$ where
$d$ is the dimension of space-time.

Though our results apply generally to the standard factorizable processes, we will continue to 
use SIDIS as a reference 
process for explaining the definitions.   
Let us also rewrite the TMD-factorization formula for SIDIS in Eq.~(\ref{eq:parton3}) as
\begin{multline}
\label{eq:SIDIS}
W^{\mu \nu} = \\
\sum_f |\mathcal{H}_f(Q;\mu)^2|^{\mu \nu} \int \, d^2 {\bf k}_{1T} \, d^2 {\bf k}_{2T} 
\delta^{(2)}({\bf k}_{1T} + {\bf q}_T - {\bf k}_{2T})\\ \times
F_{f/p}(x,{\bf k}_{1T};\mu;\zeta_F) \, D_{h/f}(z,z {\bf k}_{2T};\mu;\zeta_D) \\
= \sum_f |\mathcal{H}_f(Q;\mu)^2|^{\mu \nu} 
\int \, \frac{d^2 \trans{b}}{(2 \pi)^2} \, e^{-i {\bf q}_T \cdot \trans{b}} \\ \times \tilde{F}_{f/p}(x,\trans{b};\mu;\zeta_F) 
\, \tilde{D}_{h/f}(z,\trans{b};\mu;\zeta_D).
\end{multline}
Throughout this paper, it will be implicit that all momentum components 
are in the hadron frame.  (The hadron frame is where both hadrons 
have zero transverse momentum and is a natural frame for setting up the steps for factorization.)

Hereafter, the $Y(Q,\trans{q})$ term that appeared in Eq.~(\ref{eq:parton3}) will also be dropped because our 
primary interest is in the $q_T << Q$ regime where TMD-factorization is appropriate.  
Also, we will drop any explicit $+ \mathcal{O}((\Lambda / Q)^a)$ symbols.
We have written the TMD-factorization formula in  
coordinate space in the second equation of~(\ref{eq:SIDIS}) because it is simpler to explain the 
coordinate space definitions of the TMDs and their evolution. 
Later we will Fourier transform the TMDs back to momentum space when we analyze them numerically.

\section{Definitions of the TMDs}
\label{sec:defins}

As explained in Sect.~\ref{sec:overview}, our calculations 
are based on the formulation of TMD-factorization explained in detail
in Ref.~\cite{collins}.  A repeat of the derivation is beyond the scope of this paper.
However, in order to put our later calculations into their proper context, 
we will give an overview of the basic features of the formalism in this and 
the next section.  
We refer the reader directly to Ref.~\cite{collins} for pertinent details.

\subsection{Soft Factor Definition}
\label{sec:softfactordefinition}

We have already stressed in 
Sect.~\ref{sec:confusing}
that the definitions of the TMDs in Eq.~(\ref{eq:SIDIS}) are not the often quoted 
matrix elements of the form $\sim \langle P | \bar{\psi} \; {\rm Wilson \, Line} \; \psi | P \rangle$ with simple light-like 
Wilson lines connecting the field operators.
Using such definitions in a factorization formula 
leads to inconsistencies, including unregulated light-cone divergences.
Also, soft gluons with rapidity intermediate between 
the two nearly light-like directions
need to be accounted for in the form of soft factors.  
Therefore, before we can discuss the definitions of the TMDs
that will ultimately be used in Eq.~\eqref{eq:SIDIS}, we must provide the precise definition of the soft factor.
In coordinate space it is an expectation value of a Wilson loop:
\begin{widetext}
\begin{equation}
\label{eq:soft}
\tilde{S}_{(0)}(\trans{b};y_A,y_B) = \frac{1}{N_c} \langle 0 |W(\trans{b}/2,\infty;n_B)^{\dagger}_{ca} \, W(\trans{b}/2,\infty;n_A)_{ad} 
W(-\trans{b}/2,\infty;n_B)_{bc} W(-\trans{b}/2,\infty;n_A)^{\dagger}_{db} | 0 \rangle_{\rm No \; S.I.}.
\end{equation}
\end{widetext}
We have used the vectors in Eq.~(\ref{eq:nLLdir}) to define the directions of the Wilson 
lines so that, as long as $y_A$ and $y_B$ are finite, the Wilson lines in Eq.~(\ref{eq:soft}) are non-light-like.
The subscripts $a,b,c$ and $d$ are color triplet indices, and repeated indices are summed over.  
The ``$(0)$'' subscript indicates that bare fields are used.
The soft factor contains Wilson line self-interaction (S.I.) divergences that are very badly divergent and are unrelated to the 
original unfactorized graphs.
They must therefore be excluded, and we indicate this with 
a subscript ``${\rm No \; S.I.}$''.  
We emphasize, however, that this is only a temporary requirement because all Wilson line self-energy contributions 
will cancel in the final definitions.  
Another potential complication, pointed out in Refs.~\cite{Belitsky:2002sm,Boer:2003cm}, is that exact 
gauge invariance requires the Wilson lines to be closed by the insertion of links at light-cone infinity in 
the transverse direction.  However, the transverse segments will not contribute in the final definitions 
of the TMDs (at least in non-singular
gauges), so we do not show them explicitly in Eq.~\eqref{eq:soft}.  Again, the final arrangement of soft factors 
will ensure a cancellation.

Rather than appearing as a separate factor in the TMD-factorization formula, soft factors like Eq.~(\ref{eq:soft}) will 
be part of the final definitions of the TMDs. 
Their role in the definitions will be essential for the internal consistency of the TMDs and their validity 
in a factorization formula like Eq.~(\ref{eq:SIDIS}).

\subsection{TMD PDF and FF Definitions}
\label{sec:tmdpdfdefinition}

Now we turn to the definitions of the TMDs themselves, starting with the unpolarized TMD PDF.
The most natural first attempt at an operator definition is obtained simply by 
direct extension of the collinear integrated parton distribution, though with the Wilson line tilted to avoid
light-cone singularities.  
The operator definition is  
\begin{widetext}
\begin{multline}
\label{eq:PDFunsub}
\tmdpdf_{f/P}^{\rm unsub}(x,\trans{b};\mu;y_P - y_B) \\
= {\rm Tr}_{C} {\rm Tr}_D \int \frac{d w^{-}}{2 \pi} e^{-i x P^+ w^-} \langle P | \bar{\psi}_f(w/2) W(w/2,\infty,n_B)^\dagger 
\frac{\gamma^+}{2} W(-w/2,\infty,n_B) \psi_f(-w/2) | P \rangle_{c,\rm No \; S.I.}.
\end{multline}
\end{widetext}
This definition does not account for the overlap of the soft and collinear regions, so we refer to it as the ``unsubtracted'' 
TMD PDF.
Here $w = (0,w^-,{\bf b}_T)$ and $y_P$ is the physical rapidity of the hadron.  
As usual, the struck quark has a longitudinal plus-component of momentum $k^+ \equiv x P^+$. 
The non-light-like direction of the Wilson line is given by the $n_B$ vector defined in Eq.~(\ref{eq:nLLdir}) 
so that the light-cone divergences are regulated by the finite rapidity $y_B$.  
As usual, the definition includes a trace over color.  
The $c$ subscript is to indicate that only connected diagrams are included.
Equation~(\ref{eq:PDFunsub}) reduces exactly to the most naive definition of the TMD PDF when 
the light-like limit of $y_B \to -\infty$ is taken.
Indeed, in the final definition we will take this limit,
but then the role of the soft factors becomes important.

While the definition in Eq.~(\ref{eq:PDFunsub}) is intuitively appealing, modifications are needed 
in order to have a consistent definition that can be used in a factorization formula like Eq.~(\ref{eq:SIDIS}).
The complete definition for a quark $f$ in proton $P$, given in Refs.~\cite{collins,collins_trento}, is
\begin{widetext}
\begin{equation}
\label{eq:TMDPDFdef}
\tmdpdf_{f/P}(x,\trans{b};\mu;\zeta_F) = \tmdpdf^{\rm unsub}_{f/P}(x,\trans{b};\mu;y_P - (-\infty)) 
\sqrt{\frac{\tilde{S}_{(0)}(\trans{b};+\infty,y_s)}{\tilde{S}_{(0)}(\trans{b};+\infty,-\infty) \tilde{S}_{(0)}(\trans{b};y_s,-\infty)}} Z_F \, Z_2.
\end{equation}
Here, the ``$\infty$'' arguments for the rapidity variables in the unsubtracted PDF and the soft factors are meant in the sense of a limit. 
All field operators are unrenormalized, and $Z_F$ and $Z_2$ 
are the PDF and field strength renormalization factors respectively. The soft factors on the 
right-hand side of Eq.~(\ref{eq:TMDPDFdef}) contain rapidity arguments $y_s$.  It is an 
arbitrary parameter which can be thought of as separating the extreme plus and minus directions.
It will be convenient to express the dependence on $y_s$ via $\zeta_F$, defined in Eq.~(\ref{eq:zetaF}).
On the left-hand side of Eq.~(\ref{eq:TMDPDFdef}), the dependence on $y_s$ is expressed via the dependence on $\zeta_{F}$.

Although we will not repeat the derivation that leads 
to Eq.~(\ref{eq:TMDPDFdef}), we remark that the definition
is unique 
given the requirements that: 
a.) Factorization holds with maximal universality for the TMDs.  
b.) No explicit 
soft factor appears in the final factorization formula, Eq.~(\ref{eq:SIDIS}). 
c.) Self-interactions of the Wilson lines, and attachments to 
gauge links at infinity cancel in the final definition. 
d.) The Collins-Soper (CS) equations are homogeneous. 

There is an analogous definition for the TMD FF.  
The unsubtracted version, analogous to Eq.~(\ref{eq:PDFunsub}), is
\begin{multline}
\label{eq:FFunsub}
\tilde{D}_{H/f}^{\rm unsub}(z,\trans{b};\mu;y_A - y_h) 
= \sum_X \, \frac{1}{4 N_{c,f}} {\rm Tr}_{C} {\rm Tr}_D 
\frac{1}{z} \int \frac{d w^{-}}{2 \pi} e^{i k^+ w^-} \langle 0 | \gamma^+ \, W(w/2,\infty,n_A)
\psi_f(w/2) | h,X \rangle \\ \times \langle h, X  | \bar{\psi}_f(-w/2) W(-w/2,\infty,n_A)^\dagger | 0 \rangle_{c,\rm No \; S.I.}.
\end{multline}
Now $y_h$ is the physical rapidity of the produced hadron or jet.
The complete definition of the TMD FF with the soft factors included is:
\begin{equation}
\label{eq:TMDFFdef}
\tilde{D}_{H/f}(z,\trans{b};\mu;\zeta_D) = \tilde{D}^{\rm unsub}_{H/f}(z,\trans{b};\mu;+\infty - y_h) 
\sqrt{\frac{\tilde{S}_{(0)}(\trans{b};y_s,-\infty)}{\tilde{S}_{(0)}(\trans{b};+\infty,-\infty) \tilde{S}_{(0)}(\trans{b};+\infty,y_s)}} Z_D \, Z_2.
\end{equation}
\end{widetext}
Again there is dependence on the soft rapidity $y_s$.
For the FF, the energy cutoff scale $\zeta_D$ is related to the soft rapidity scale $y_s$ via Eq.~\eqref{eq:zetaD}.

Equations~(\ref{eq:TMDPDFdef}) and~(\ref{eq:TMDFFdef}) are the correct  
TMDs for the TMD-factorization formula in Eq.~(\ref{eq:SIDIS}) as well as for $e^+ e^-$ annihilation with back-to-back jets.
Up to a flip in the direction of the Wilson line from future to past pointing, which is important for accounting for a sign flip in 
certain types of TMDs, the TMD PDF in~(\ref{eq:TMDPDFdef}) is also relevant for the Drell-Yan process.    
In this section we have clarified the meaning of the energy parameters $\zeta_F$ and $\zeta_D$, which were already discussed in 
Sects.~\ref{sec:overview},~\ref{sec:setupandnotation}.
Together, Eqs.~(\ref{eq:zetaF},~\ref{eq:zetaD}) give $\sqrt{\zeta_F \zeta_D} \approx Q^2$.
They are in principle arbitrary, and the full factorization 
formula in Eq.~(\ref{eq:SIDIS}) is exactly independent of the choice of $y_s$ (and therefore $\sqrt{\zeta_F}$ and $\sqrt{\zeta_D}$).
However, different choices are needed for each factor in Eq.~\eqref{eq:SIDIS} in order to optimize the 
convergence properties of the perturbation series.
To obtain the TMDs appropriate for different scales, we must appeal to 
evolution equations which are the subject of Sect.~\ref{sec:evolved}.

\subsection{The Role of Soft Factors}
\label{sec:intuitive}

Admittedly, the final definitions in Eqs.~\eqref{eq:TMDPDFdef} and~\eqref{eq:TMDFFdef} appear rather complex.
While detailed derivations are beyond the scope of this paper, it is 
nevertheless worthwhile to make 
a
few intuitive remarks about how these definitions
arise in a treatment of factorization.  A much more detailed treatment is found in Ref.~\cite{collins}.
  
For now we simplify the notation for the TMDs by dropping all arguments and symbols not directly 
related to Wilson line rapidities.   The  
cross section can 
then be written (schematically) as,
\begin{equation}
\label{eq:factschem}
d \sigma = |\mathcal{H}|^2 \, \frac{\tmdpdf^{\rm unsub}(y_P - (-\infty)) \times \tilde{D}^{\rm unsub}(+\infty - y_h)}{\tilde{S}(+\infty,-\infty)}.
\end{equation}
The $\tmdpdf^{\rm unsub}(y_P - (-\infty))$ and $\tilde{D}^{\rm unsub}(+\infty - y_h)$ are the same ``unsubtracted'' TMDs 
from Eqs.~\eqref{eq:PDFunsub} and~\eqref{eq:FFunsub}.  They each describe the distribution of 
gluons in their relevant collinear direction, but they also both account for soft gluons with nearly zero rapidity.
Therefore, the soft factor $\tilde{S}(+\infty,-\infty)$ in the denominator is needed to remove double counting.  
Since the Wilson lines in the unsubtracted TMDs are light-like in the plus and minus directions respectively,
they also include rapidity divergences.  Thus, the role of the $\tilde{S}(+\infty,-\infty)$ is also 
to cancel these rapidity divergences.  Although Eq.~\eqref{eq:factschem} properly accounts for all soft and 
collinear regions and deals with the divergences, it  
is not factorized.  Because of 
the rapidity divergences, it is immediately clear that 
$\tmdpdf^{\rm unsub}(y_P - (-\infty))$ and $\tilde{D}^{\rm unsub}(+\infty - y_h)$ are not separately 
well-defined, and in the full formula they are entangled via the soft 
denominator.  
To get a factorized structure for Eq.~\eqref{eq:factschem}, with each factor individually well-defined,
a natural first step to try is 
simply to separate the 
soft factor into a product of two factors:
\begin{equation}
\label{eq:factschem2}
d \sigma = |\mathcal{H}|^2 \,
\frac{\tmdpdf^{\rm unsub}(y_P - (-\infty))}{\sqrt{\tilde{S}(+\infty,-\infty)}} \times 
\frac{\tilde{D}^{\rm unsub}(+\infty - y_h)}{\sqrt{\tilde{S}(+\infty,-\infty)}}.
\end{equation}  
One is then tempted to identify the factors 
on
either side of the ``$\times$'' 
with the TMD PDF and the TMD FF.  However, these definitions still contain 
uncanceled divergences.  In each factor, the rapidity divergence in the numerator is 
not completely canceled by the square root rapidity divergence in the denominator, and
new rapidity divergences are introduced by the Wilson line 
pointing in the opposite direction.  
So the next modification of Eq.~\eqref{eq:factschem} is to write
\begin{equation}
\label{eq:factschem3}
\begin{split}
d \sigma & =  |\mathcal{H}|^2 \,
\frac{\tmdpdf^{\rm unsub}(y_P - (-\infty))}{\sqrt{\tilde{S}(+\infty,-\infty)}} \times \\ 
 \times & 
\frac{\sqrt{\tilde{S}(+\infty,-\infty)}}{\sqrt{\tilde{S}(+\infty,-\infty)}} \times
\frac{\tilde{D}^{\rm unsub}(+\infty - y_h)}{\sqrt{\tilde{S}(+\infty,-\infty)}} \\ 
& =  |\mathcal{H}|^2 \, \frac{\tmdpdf^{\rm unsub}(y_P - (-\infty))}{\sqrt{\tilde{S}(+\infty,-\infty)}} \times \\ 
 \times &
\frac{\sqrt{\tilde{S}(+\infty,y_s) \, \tilde{S}(y_s,-\infty)}}{\sqrt{\tilde{S}(+\infty,y_s) \, \tilde{S}(y_s,-\infty)}} \times
\frac{\tilde{D}^{\rm unsub}(+\infty - y_h)}{\sqrt{\tilde{S}(+\infty,-\infty)}} \\
& = 
|\mathcal{H}|^2 \times \\ 
 \times & 
\left\{ \tmdpdf^{\rm unsub}(y_P - (-\infty)) \sqrt{\frac{\tilde{S}(+\infty,y_s)}{\tilde{S}(+\infty,-\infty)  \tilde{S}(y_s,-\infty)}}  \right\} 
\times \\ 
 \times & 
\left\{ \tilde{D}^{\rm unsub}(+\infty - y_h) \sqrt{\frac{\tilde{S}(y_s,-\infty)}{\tilde{S}(+\infty,-\infty) \tilde{S}(+\infty,y_s)}} \right\}.
\end{split}
\end{equation}  
After the first equality, we have simply multiplied and divided by $\sqrt{\tilde{S}(+\infty,-\infty)}$.
After the second equality, we have used the group relation $\tilde{S}(y_A,y_C) \propto \tilde{S}(y_A,y_B) \tilde{S}(y_B,y_C)$, 
which follows from the evolution equations for the soft factor --- see, e.g., Ref.~\cite{collins} chapter 10.
(In fact, this expression should also include an overall factor that depends on rapidity $y_B$.  But this 
cancels in Eq.~\eqref{eq:factschem3} between the numerator and denominator and does not affect our argument.)
This allows for a separation of the soft factors into pieces that have rapidity 
divergences only in the plus or only in the minus directions, 
with any other rapidity divergences cut off by the arbitrary scale $y_s$.  By rearranging the soft factors,
everything can then be grouped into the factors on the last two lines.
In each factor, all spurious divergences cancel, and we arrive at the separately well-defined TMDs in braces.
These correspond to the definitions in Eqs.~\eqref{eq:TMDPDFdef} and~\eqref{eq:TMDFFdef}.

To summarize, the light-like Wilson lines are needed in each 
separate ``unsubtracted'' TMD of Eq.~\eqref{eq:factschem}, but the 
contribution from 
gluon attachments to a Wilson line where the 
gluon has nearly the \emph{same rapidity} as the Wilson line does not correspond 
to any real physics.  To cancel these spurious contributions to the cross section, there must be an equal 
number of both plus-pointing and minus-pointing
light-like Wilson lines in the numerator and denominator, as is the case in Eq.~\eqref{eq:factschem}.
Applying this same requirement to the separate TMDs (a TMD PDF and a TMD FF, in our case) leads uniquely 
to the definition in the last two lines of Eq.~\eqref{eq:factschem3} and Eqs.~\eqref{eq:TMDPDFdef} 
and~\eqref{eq:TMDFFdef}.  Compare this with the situation in Ref.~\cite{Becher:2010tm}.  There, as in our Eq.~\eqref{eq:factschem},
the needed cancellations occur in the full cross section expression, but not in the individual TMD factors. 
To get separately consistent TMDs, the steps summarized in Eq.~\eqref{eq:factschem3} are needed.

\section{Evolved TMDs}
\label{sec:evolved}

The evolution of the TMDs follows from their definitions, 
Eqs.~(\ref{eq:TMDPDFdef},~\ref{eq:TMDFFdef}). 
We start with the evolution of the TMD PDF.
The energy evolution is given by the  
CS-equation for Eq.~(\ref{eq:TMDPDFdef}):
\begin{equation}
\label{eq:CSPDF}
\frac{\partial \ln \tilde{F}(x,\trans{b};\mu,\zeta_F)}{\partial \ln \sqrt{\zeta_F} } = \tilde{K}(\trans{b};\mu) 
\end{equation}
where the function $\tilde{K}(\trans{b};\mu)$ is defined as,
\begin{equation}
\label{eq:KPDF}
\tilde{K}(\trans{b};\mu) = 
\frac{1}{2} \frac{\partial}{\partial y_s} \ln \left( \frac{\tilde{S}(\trans{b};y_s,-\infty)}{\tilde{S}(\trans{b};+\infty,y_s)} \right).
\end{equation}
Equation~(\ref{eq:CSPDF}) follows directly from differentiating  
Eq.~(\ref{eq:TMDPDFdef}) with respect to $\sqrt{\zeta_F}$ and using the definition of $\tilde{K}(\trans{b};\mu)$.
Note that it is $\tilde{S}(\trans{b})$ rather than $\tilde{S}_{(0)}(\trans{b})$ that appears in Eq.~\eqref{eq:KPDF}.  Thus  
it is important to account for the UV renormalization factors $Z_F Z_2$ in Eq.~\eqref{eq:TMDPDFdef}.

The RG equations for both $\tilde{F}(x,\trans{b};\mu;\zeta_F)$ and $\tilde{K}(\trans{b};\mu)$ are also needed. 
They are,
\begin{equation}
\label{eq:RGKPDF}
\frac{d \tilde{K}(\trans{b};\mu)}{d \ln \mu} = - \gamma_K(g(\mu))
\end{equation}
and
\begin{equation}
\label{eq:RGPDF}
\frac{d \ln \tilde{F}(x,\trans{b};\mu,\zeta_F)}{d \ln \mu} = \gamma_F(g(\mu);\zeta_F /\mu^2).
\end{equation}
The functions $\gamma_K(g(\mu))$ and $\gamma_F(g(\mu);\zeta_F /\mu^2)$ are the anomalous dimensions of 
$\tilde{K}(\trans{b};\mu)$ and $\tilde{F}(x,\trans{b};\mu,\zeta_F)$ respectively.
Using Eqs.~(\ref{eq:CSPDF}-\ref{eq:RGPDF}), the energy evolution of $\gamma_F$ can be derived:
\begin{multline}
\label{gammaF}
\gamma_F(g(\mu);\zeta_F /\mu^2) = \gamma_F(g(\mu);1) - \frac{1}{2} \gamma_K(g(\mu)) \ln \frac{\zeta_F}{\mu^2}.
\end{multline}

At small-$b_T$, Eq.~(\ref{eq:TMDPDFdef}) can itself be calculated within a collinear 
factorization formalism~\cite{CSS1}.  Namely, it separates into a perturbatively calculable 
hard scattering coefficient and an integrated PDF, convoluted over momentum fraction:
\begin{multline}
\label{eq:smallb}
\tilde{F}_{f/P}(x,\trans{b};\mu,\zeta_F) = \\
=\sum_j\int_x^1\frac{d\hat{x}}{\hat{x}}  \tilde{C}_{f/j}(x/\hat{x},b_T;\zeta_F,\mu,g(\mu)) f_{j/P}(\hat{x};\mu) \\
+ \mathcal{O}((\Lambda_{\rm QCD} b_T)^a).
\end{multline}
The functions $f_{j/P}(\hat{x};\mu)$ are the ordinary integrated PDFs 
and the $\tilde{C}_{f/j}(x/\hat{x},b_T;\zeta_F,\mu,g(\mu))$
are the hard coefficient functions, which are provided  
to first order in Appendix~\ref{sec:coefffuncts}.
The last term denotes the error, which grows large when $b_T \gtrsim \Lambda_{\rm QCD}^{-1}$.

At large $b_T$, the perturbative treatment of the $b_T$-dependence is no longer 
reliable.  In momentum space, this corresponds to the breakdown of the
perturbative treatment of the $k_T$-dependence at small-$k_T$.
It is in this region that the concept of TMD-factorization, 
incorporating TMDs with intrinsic non-perturbative transverse momentum, becomes very important.

While the $b_T$-dependence at large $b_T$ cannot be calculated directly in pQCD,
the scale dependence can still be handled with the evolution equations~(\ref{eq:CSPDF}-\ref{gammaF}).
But a prescription is needed for matching the large and small $b_T$-behavior.
The most common matching procedure was developed in 
Ref.~\cite{Collins:1981va}.  It replaces $\trans{b}$ in the hard part of the calculation by
a function,
\begin{equation}
\label{bstar}
{\bf b}_{\ast}(\trans{b}) \equiv \frac{\trans{b}}{\sqrt{1+b_T^2/b_{\rm max}^2}}.
\end{equation}
This definition of ${\bf b}_{\ast}(\trans{b})$ is constructed so that 
it is equal to $\trans{b}$ when $\trans{b}$ is small, while smoothly approaching an upper cutoff $b_{\rm max}$ when $\trans{b}$
becomes too large.
The value of $b_{\rm max}$ is typically chosen to be of order 
$\sim 1$ GeV$^{-1}$ and should be thought of as characterizing the boundary of the 
perturbative region of the ${\bf b}_T-dependence$.

In the calculation of the hard coefficient in Eq.~(\ref{eq:smallb}), 
the appropriate size for the scale $\mu$ is determined by the size of ${\bf b}_{\ast}(\trans{b})$.
Hence, we define the scale, 
\begin{equation}
\label{eq:mub}
\mu_b = \frac{C_1}{b_{\ast}(\trans{b})}. 
\end{equation}
The parameter $C_1$ is chosen to optimize the perturbation expansion.
For all our calculations, we will use $C_1 = 2 e^{-\gammae}$.
At large $b_T$ in the final 
expression for the evolved TMD PDF, the effect of the deviation between $\trans{b}$ and ${\bf b}_{\ast}$ in 
$\tmdpdf_{f/P}(x,\trans{b};\mu;\zeta_F)$ and $\tilde{K}(\trans{b};\mu)$ will be accounted for by 
extra non-perturbative, but universal and scale-independent, functions.

Applying the evolution equations in Eqs.~(\ref{eq:CSPDF}-\ref{gammaF}), using the collinear 
factorization treatment for small $\trans{b}$ from Eq.~(\ref{eq:smallb}), 
and implementing the matching procedure of Eq.~(\ref{bstar}) allows the TMD PDF to be written with maximum 
perturbative input in terms of evolution 
from fixed starting scales: 
\begin{widetext}
\begin{multline}
\label{eq:evolvedPDF}
\tilde{F}_{f/P}(x,\trans{b};\mu,\zeta_F) = 
\stackrel{\rm A}{\overbrace{\sum_j \int_x^1 \frac{d \hat{x}}{\hat{x}} \tilde{C}_{f/j}(x/\hat{x},b_{\ast};\mu_b^2,\mu_b,g(\mu_b)) f_{j/P}(\hat{x},\mu_b)}} \\
\stackrel{\rm B}{\overbrace{\times \exp \left\{ \ln \frac{\sqrt{\zeta_F}}{\mu_b} \tilde{K}(b_{\ast};\mu_b) + 
\int_{\mu_b}^\mu \frac{d \mu^\prime}{\mu^\prime} \left[ \gamma_F(g(\mu^\prime);1) 
- \ln \frac{\sqrt{\zeta_F}}{\mu^\prime} \gamma_K(g(\mu^\prime)) \right]\right\}}} 
\times 
\stackrel{\rm C}{\overbrace{\exp \left\{ g_{j/P}(x,b_T) + g_K(b_T) \ln \frac{\sqrt{\zeta_F}}{\sqrt{\zeta_{F,0}}} \right\}}}.
\end{multline}
This is our master equation for fitting TMD PDFs while incorporating evolution.
For a much more detailed explanation of the steps summarized above and 
leading to Eq.~(\ref{eq:evolvedPDF}), we again refer the reader to Ref.~\cite{collins}, especially chapters 10 and 13.
The steps for evolving are very similar to traditional applications of the CSS formalism, but now 
they are applied to separate, individual TMDs.
The scales used in the evolution are chosen to minimize the size of higher order corrections in 
the perturbatively calculable parts.
We have labeled three separate factors by ``A'', ``B'', and ``C'' to aid in the detailed discussion 
that will appear in the next section.  
The $\tilde{C}_{f/j}(x/\hat{x},b_{\ast};\mu_b^2,\mu_b,g(\mu_b))$, $\tilde{K}(b_{\ast};\mu_b)$, $\gamma_F(g(\mu^\prime);1)$, 
and $\gamma_K(g(\mu^\prime))$ functions are all perturbatively calculable for all $\trans{b}$.
They are provided to order $\alpha_s$ in Appendices.~\ref{sec:coefffuncts} and~\ref{sec:anomdims}.
On the first line, $f_{j/P}(\hat{x},\mu_b)$ is the ordinary 
integrated PDF from collinear factorization.  
The functions $g_{j/P}(x,b_T)$ and $g_K(b_T)$ describe
the non-perturbative $\trans{b}$-behavior in $\tmdpdf_{f/P}(x,\trans{b};\mu;\zeta_F)$
and $\tilde{K}(\trans{b};\mu)$ respectively.  They are scale-independent and universal.  The function $g_K(b_T)$ is notably 
independent of the species of external hadrons.
Our definition of the factor $g_{j/P}(x,b_T)$ differs slightly from what 
is used in~\cite{collins} because it has absorbed a term equal to $g_K(b_T) \ln (\sqrt{\zeta_{F,0}} / x M_p)$.
This will allow us to choose an arbitrary starting scale $\zeta_{F,0}$ for the evolution in $\sqrt{\zeta_F}$. 

We are ultimately interested in the momentum space TMD which is just the 
Fourier transform of the coordinate space TMD PDF in Eqs.~(\ref{eq:TMDPDFdef},\ref{eq:evolvedPDF}):
\begin{equation}
\label{eq:TMDPDFdefFT}
F_{f/P}(x,\trans{k};\mu,\zeta_F) =
\frac{1}{(2 \pi)^2} \int d^2 {\bf b}_T \, e^{i {\bf k}_T \cdot {\bf b}_T} \, \tmdpdf_{f/P}(x,\trans{b};\mu,\zeta_F).
\end{equation}
Once Eq.~(\ref{eq:evolvedPDF}) has been parametrized, $F_{f/P}(x,\trans{k};\mu,\zeta_F)$ can be determined
directly by a numerical Fourier transform.  

Exactly analogous steps hold for the TMD FF.  
It is related to $\tilde{K}(\trans{b};\mu)$ by,
\begin{equation}
\label{eq:CSFF}
\frac{\partial \ln \tilde{D}(z,\trans{b};\mu,\zeta_D)}{\partial \ln \sqrt{\zeta_D} } = \tilde{K}(\trans{b};\mu). 
\end{equation}
There is also an RG equation analogous to Eq.~(\ref{eq:RGPDF}), with anomalous dimension $\gamma_D(g(\mu);\zeta_D /\mu^2)$:
\begin{equation}
\label{eq:RGFF}
\frac{d \ln \tilde{D}(z,\trans{b};\mu,\zeta_D)}{d \ln \mu} = \gamma_D(g(\mu);\zeta_D /\mu^2).
\end{equation}
For the small-$\trans{b}$ region, the 
collinear factorization treatment of Eq.~(\ref{eq:TMDFFdef}), analogous to Eq.~(\ref{eq:smallb}), gives
\begin{equation}
\label{eq:smallbFF}
\tilde{D}_{H/f}(z,\trans{b};\mu,\zeta_D) = 
\sum_j\int_z^1\frac{d\hat{z}}{\hat{z}^{3 - 2 \epsilon}}  \tilde{C}_{j/f}(z/\hat{z},b_T;\zeta_D,\mu,g(\mu)) d_{h/j}(\hat{z};\mu) \\
+ \mathcal{O}((\Lambda_{\rm QCD} b_T)^a)
\end{equation}
The analogue of Eq.~(\ref{eq:evolvedPDF}) for the TMD FF is
\begin{multline}
\label{eq:evolvedFF}
\tilde{D}_{H/f}(z,\trans{b};\mu,\zeta_D) = 
\sum_j \int_z^1 \frac{d \hat{z}}{\hat{z}^{3-2\epsilon}} \tilde{C}_{j/f}(z/\hat{z},b_{\ast};\mu_b^2,\mu_b,g(\mu_b)) d_{H/j}(\hat{z},\mu_b) \\
\times \exp \left\{ \ln \frac{\sqrt{\zeta_D}}{\mu_b} \tilde{K}(b_{\ast};\mu_b) + 
\int_{\mu_b}^\mu \frac{d \mu^\prime}{\mu^\prime} \left[ \gamma_D(g(\mu^\prime);1) 
- \ln \frac{\sqrt{\zeta_D}}{\mu^\prime} \gamma_K(g(\mu^\prime)) \right]\right\} 
\times 
\exp \left\{ g_{h/j}(z,b_T) + g_K(b_T) \ln \frac{\sqrt{\zeta_D}}{\sqrt{\zeta_{D,0}}} \right\}.
\end{multline}
\end{widetext}
As with the TMD PDF, the perturbative parts of Eq.~(\ref{eq:evolvedPDF}) have been calculated to order $\alpha_s$ and 
are supplied for reference in the appendices.  (Note the factor of $\hat{z}^{2\epsilon - 3}$ that appears in the 
integration measure in Eq.~(\ref{eq:evolvedFF}) as compared to the $\hat{x}^{-1}$ factor that appears in Eq.~(\ref{eq:evolvedPDF}); this is 
due to differences in normalization of the integrated PDFs and FFs).
The non-perturbative function $g_K(b_T)$ is the same in both the TMD PDF and FF.
The function $g_{h/j}(z,b_T)$ describes the non-perturbative large-$b_T$ behavior that is 
specific to a fragmentation function for parton $j$ and 
hadron $H$.
The momentum-space TMD FF is defined to be
\begin{multline}
\label{eq:TMDFFdefFT}
D_{H/f}(z,z \trans{k};\mu,\zeta_D) = \\
\frac{1}{(2 \pi)^2} \int d^2 {\bf b}_T \, e^{-i {\bf k}_T \cdot {\bf b}_T} \, \tilde{D}_{H/f}(z,\trans{b};\mu,\zeta_D).
\end{multline}
Note that the standard momentum space definition has $z \trans{k}$ as the transverse momentum argument rather than $\trans{k}$.

The important result of this section is that we now have expressions for the evolved TMD quark PDF and FF
that can be used in Eq.~(\ref{eq:parton3}), which in turn has a very similar structure to the generalized parton 
model picture in Eq.~(\ref{eq:parton1}).

\section{Implementing Evolution}
\label{sec:impevol}
We now discuss explicit calculations of evolved momentum space TMDs.   
Given some non-perturbative input for the large-$b_T$ behavior 
at some fixed scales, we can calculate the TMDs at different scales 
by directly calculating Eqs.~(\ref{eq:evolvedPDF}--\ref{eq:evolvedFF}).
We will discuss the TMD PDF and FF cases separately.

\subsection{TMD PDFs}
\label{sec:pdfevolv}

We first analyze the TMD PDF by discussing each factor labeled in Eq.~(\ref{eq:evolvedPDF}) separately.
The first factor (the A-factor) matches the TMD PDF to a collinear 
treatment in the small $b_T  << 1 / \Lambda_{QCD}$ limit.
As with standard collinear factorization, it involves a hard part, which in this case is the coefficient function 
$\tilde{C}_{f/j}$, and a collinear factor, which is just the standard integrated PDF. 
At lowest order in a calculation of the coefficient function, the A-factor is 
simply $f(x,\mu_b)$.  
The first factor on the second line, the B-factor, is an exponential of quantities that can all be 
calculated perturbatively.  They are the CS kernel $\tilde{K}(b_{\ast};\mu_b)$ at small $b_T$, 
the anomalous dimension $\gamma_F$ of the 
TMD PDF, and the 
anomalous dimension $\gamma_K$ of the CS kernel.
The last factor, the C-factor, implements the 
matching between the small and large $b_T$-dependence.
The function $g_{j/P}(x,b_T)$ 
parametrizes the non-perturbative large-$b_T$ behavior that is intrinsic to the proton, 
while $g_K(b_T)$ parametrizes the non-perturbative 
large-$\trans{b}$ behavior of $\tilde{K}(\trans{b};\mu)$.
The function $g_{j/P}(x,b_T)$ is universal, but in principle depends on the external hadron.
The function $g_K(b_T)$ is both universal and independent of the species of external hadrons. 
Note that, while the description of the $b_T$-behavior becomes non-perturbative at large $b_T$, there
is still perturbatively calculable evolution 
for the TMD
coming from the B-factor.

For doing calculations, a choice for the numerical values of 
$\zeta_F$ and $\zeta_D$ in Eqs.~(\ref{eq:evolvedPDF},\ref{eq:evolvedFF})
is needed.  Since $\sqrt{\zeta_F \zeta_D} \approx Q^2$, we will 
treat the PDFs and FFs symmetrically and use $\sqrt{\zeta_F} = \sqrt{\zeta_D} = Q$.
(In principle, slightly different choices may be preferred in specific applications, but this will be sufficient for now.)
Also, we relabel $\sqrt{\zeta_{F,0}} = \sqrt{\zeta_{D,0}} \equiv 2 Q_0$.

\begin{figure*}
\centering
\includegraphics[scale=.5]{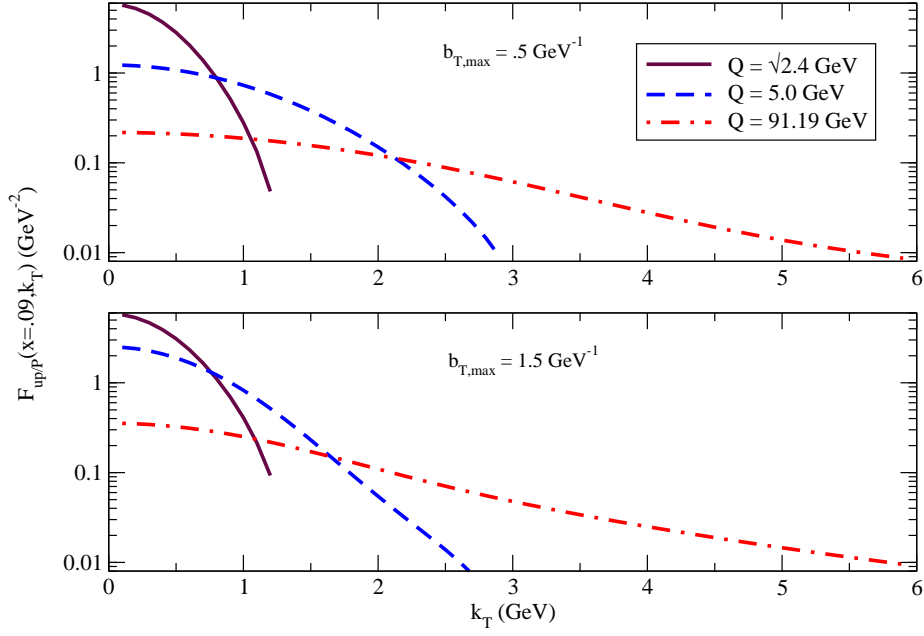}
\caption{The up quark TMD PDF for $Q = \sqrt{2.4},5.0$ and $91.19$~GeV and $x = 0.09$. The upper plot shows the result 
of using the BLNY fit in Eq.~(\ref{eq:BLNY3}) with $b_{max} = 0.5$~GeV$^{-1}$ while the lower panel shows the BLNY fit 
obtained with  $b_{max} = 1.5$~GeV$^{-1}$.
The solid maroon, dashed blue, and red dot-dashed curves are for $Q = \sqrt{2.4},5.0$ and $91.19$~GeV respectively (see online version 
for color).
}
\label{fig:TMDPDF}
\end{figure*}

It is instructive to investigate the relationship between the 
parton model expectation and Eq.~(\ref{eq:evolvedPDF}).  
In standard collinear factorization for processes 
integrated over transverse momentum, the parton model 
description of the \emph{integrated} PDF is recovered by
dropping all order-$\alpha_s$ contributions to the DGLAP evolution 
kernel, reproducing the Bjorken scaling property of the parton model.  
In collinear factorization, the parton model can be understood as the zeroth order contribution 
to the full pQCD factorization result. 
In the TMD PDF case, however, if all order-$\alpha_s$ or higher contributions to Eq.~(\ref{eq:evolvedPDF})
are dropped, then the TMD PDF becomes:
\begin{multline}
\label{eq:zerothPDF}
\tilde{F}_{f/P}(x,\trans{b};\zeta_F,\mu) \to f_{j/P}(x) \\
\times \exp \left\{ g_{j/P}(x,b_T) + g_K(b_T) \ln \frac{ Q }{ 2 Q_0 } \right\}.
\end{multline}
Usually, a Gaussian model is used in a partonic description of the TMD PDF like Eq.~(\ref{eq:parton1}).
So we write $g_{j/P}(x,b_T)$ as  $-g_1 b_T^2/2$
and $g_K(b_T)$ as $-g_2 b_T^2/2$.  Then Eq.~(\ref{eq:zerothPDF}) becomes
\begin{equation}
\label{eq:zerothPDF2}
f_{j/P}(x)\times \exp \left\{ - \left[ g_1 + g_2 \ln \frac{Q}{ 2 Q_0 } \right] \frac{b_T^2}{2} \right\}.
\end{equation}  
This is almost the Gaussian/parton model form of the TMD PDF.
However, there is still scale dependence coming from the coefficient of the $g_K(b_T)$ function.
This difference from the collinear case is due 
to the fact that, while 
the DGLAP evolution kernels vanish when order-$\alpha_s$ terms are neglected, the evolution kernel in Eq.(\ref{eq:KPDF}) 
is non-vanishing at zeroth order because of the non-perturbative contribution at large $b_T$.
TMD-factorization therefore differs in a significant qualitative way from collinear factorization 
in that the naive expectation from the parton model picture is not exactly recovered even in a zeroth order treatment --- there is 
still potentially large scale dependence at large $b_T$.
This can have a large effect on the small-$k_T$ scale dependence of the TMDs, 
as already noted in Ref.~\cite{Boer:2001he,Boer:2008fr}.
In particular, if $g_1 \ll g_2$, then it can be seen from Eq.~(\ref{eq:zerothPDF2})
that the TMD PDF becomes extremely sensitive 
to $Q$ near $Q \sim 2 Q_0$ and at large $b_T$.  
In the momentum space TMD PDF,  
the evolution
corresponds 
to rapid suppression at small $k_T$, of order $k_T\sim 1$ GeV,
with increasing $Q$. 
The effect can be  
observed in the small-$k_T$ region of the curves in Fig.~\ref{fig:TMDPDF}.

Once the A and B and C factors are known, 
it becomes straightforward to calculate the Fourier transform in 
Eq.~(\ref{eq:TMDPDFdefFT}).
Of these, the A-factor is the most cumbersome to deal with 
because it requires numerical integrals over $x$ that involve integrated 
PDFs.  The integrated PDFs themselves need to be imported from previous fits.
In our calculations, we obtain the A-factor in Eq.~(\ref{eq:evolvedPDF}) by 
using the MSTW PDFs~\cite{Martin:2009iq}, along with 
the $\overline{{\rm MS}}$ coefficient functions calculated in Appendix~\ref{sec:coefffuncts}.
To facilitate future calculations, we have made separate tables for the A-factor available for each 
quark flavor~\cite{webpage}.  The B-factor, up to order $\alpha_s$, is straightforward to calculate directly
using the anomalous dimensions provided in Appendix~\ref{sec:anomdims}.  

All that is then needed to obtain
Eq.~(\ref{eq:TMDPDFdefFT}) is a model or a 
fit of the non-perturbative $\trans{b}$-behavior of the C-factor.
For our calculations, we appeal to currently available fits.
In principle, fitting the non-perturbative parts, $g_{j/P}(x,b_T)$ and $g_K(b_T)$, requires knowledge 
of the complete $(x,\trans{b})$ plane at different values of $Q$ and for each flavor.
There have been extensive efforts over the past several decades to determine 
these parameters from experiments, most commonly from fits to DY processes. 
Currently, the most detailed global fits use the 
Brock-Landry-Nadolsky-Yuan (BLNY) form for the full 
non-perturbative $\trans{b}$-dependence, which leads to a factor in the full cross section equal to~\cite{Landry:2002ix}:
\begin{equation}
\label{eq:BLNY}
\exp \left\{- \left[g_1 + g_2 \ln \frac{Q}{2 Q_0} + g_1 g_3 \ln(100 x_A x_B) \right] b_T^2 \right\}.
\end{equation}
The variables $x_A$ and $x_B$ are the usual 
momentum fraction variables of the annihilating quark and anti-quark.
This almost gives the simple form in Eq.~(\ref{eq:zerothPDF2}), but now there is a
term in the exponent with explicit $x$-dependence.   
In the $p \bar{p}$ cross section, two C-factors appear;
one with a function $g_{j/P}(x,b_T)$ for the probability of finding a quark in a proton, and the other with a function
$g_{\bar{j}/\bar{P}}(x,b_T)$ for finding an anti-quark in an anti-proton.
Assuming flavor independence, the symmetric role of the PDFs in the DY factorization formula allows for an immediate 
identification of the C-factor contribution to the TMD PDF in Eq.~(\ref{eq:evolvedPDF}):
\begin{equation}
\label{eq:BLNY2}
\exp \left\{- \left[\frac{g_2}{2} \ln \frac{Q}{2 Q_0} + g_1 \left( \frac{1}{2} + g_3 \ln(10 x) \right) \right] b_T^2 \right\}.
\end{equation}
The fits of Ref.~\cite{Landry:2002ix} found 
$g_1 = 0.21$~GeV$^{2}$, $g_2 = 0.68$~GeV$^{2}$ and $g_3 = -0.6$, using $Q_0 = 1.6$~GeV 
using data from Refs.~\cite{DY1,DY2,DY3,DY4,DY5,DY6}.
However, these fits mix data for $p \bar{p}$, $pp$ and $p Cu$ scattering
which means that it must be assumed that the non-perturbative functions $g_{j/P}(x,b_T)$, $g_{\bar{j}/P}(x,b_T)$   
and $g_{j/Cu}(x,b_T)$ are similar.  
This is not  
a serious problem at large $Q$ because then the $Q$-behavior 
comes mainly from the $g_K(b_T)$ function which is independent of external hadrons.
However, we also want our TMD PDF to be valid at smaller $Q \lesssim Q_0$ scales, relevant 
to many SIDIS experiments. 
Our strategy then is to match the BLNY fit
to the recent scale-independent Gaussian fits by 
Schweitzer, Teckentrup and Metz (STM)~\cite{Schweitzer:2010tt}. 
Using HERMES SIDIS data~\cite{Airapetian:2009jy,Airapetian:2002mf} 
for $\langle x \rangle = 0.09$, $\langle Q^2 \rangle \approx 2.4$~GeV$^2$ and $z > 0.2$ they find
\begin{equation}
\label{eq:STM1}
F_{f/P}(x,\trans{k}) = f_{f/P}(x) \times \frac{\exp \left[ -k_T^2 / \langle k_T^2\rangle \right]}{\pi \langle k_T^2\rangle} 
\end{equation}
with $\langle k_T^2\rangle = (.38 \pm 0.06)$~GeV$^{2}$. 
To recover this in our fit, 
we modify the BLNY parametrization in Eq.~(\ref{eq:BLNY2}) by rewriting it as,
\begin{multline}
\label{eq:BLNY3}
\exp \left\{- \left[\frac{g_2}{2} \ln \frac{Q}{2 Q_0} + \right. \right. \\
\left. \left. g_1 \left( \frac{1}{2} + g_3 \ln \left( 10 \frac{x x_0}{x_0 + x} \right) \right) \right] b_T^2 \right\}.
\end{multline} 
If $x_0 \approx 0.02$, then Eq.~(\ref{eq:BLNY3}) approximately matches the STM fit for 
$x = 0.09$ and $Q = \sqrt{2.4}$~GeV, 
but reduces to the BLNY fit at larger $Q$ and smaller $x$.
We note that the $x$ and ${\bf b}_T$ dependence does not 
quite factorize in these TMD fits.  Indeed, the form of the $g_{j/P}(x,b_T)$ 
is not required by the formalism to factorize into separate $x$ and ${\bf b}_T$ dependence.

We now have a fit that includes the
scale dependence of the QCD evolution in Eq.~(\ref{eq:evolvedPDF}), 
and whose $b_T$-dependence matches two previously 
performed fits for different regions of kinematics.
For illustration, we have plotted in Fig.~\ref{fig:TMDPDF} the TMD PDF of the up-quark 
for the small, medium and large 
values of $Q =\sqrt{2.4}$, $5$, and $91.19$~GeV and with $x = 0.09$.  
We have made the plot run over a range from $k_T = 0$ to $6$~GeV, typical for studies of TMD-functions.
(Recall, however, that without the $Y$-term of Eq.~(\ref{eq:parton3}) the TMD PDF by itself only has a simple interpretation 
for $k_T << Q$.)
Comparing the curves, it is clear that the evolution in $Q$ is a large effect, leading to more than an 
order of magnitude of suppression at small $k_T$, and a broad tail
at larger $k_T$.  Numerical computations that produce plots like Fig.~\ref{fig:TMDPDF} are available at Ref.~\cite{webpage}.
%%%%%%%%%%%%%%%%%%%%%%%%%%%%%
%%%%%%%%%%%%%%%%%%%%%%%%%%%%%
%%%%%%%%%%%%%%%%%%%%%%%%%%%%%
\begin{figure*}
\centering
\includegraphics[scale=.5]{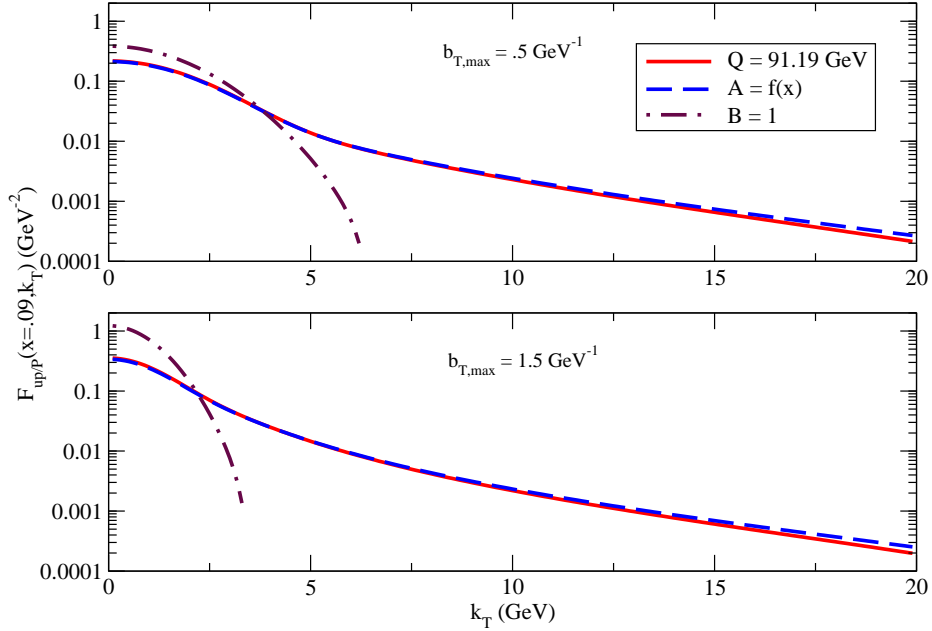}
\caption{Comparing the shape of the TMD PDF within various approximations.  The solid red curves 
are the same as the $Q = 91.19$~GeV curves in Fig.~\ref{fig:TMDPDF}.  The dashed blue curve is the result of setting 
the A-factor in Eq.~(\ref{eq:evolvedPDF}) equal to $f(x,\mu_b)$, and the dash-dotted maroon curve 
is obtained by setting the B-factor in Eq.~(\ref{eq:evolvedPDF}) equal to $1$.
(See online version for color.)}
\label{fig:PDFnoAB}
\end{figure*}
%%%%%%%%%%%%%%%%%%%%%%%%%%%%%
The large $b_T$ cutoff, $b_{\rm max}$, should be small enough to exclude the non-perturbative 
large $b_T$-regime from the perturbative part of the TMD PDF, but it is otherwise arbitrary.
In fits, different choices of $b_{\rm max}$ can lead to 
very different values for the non-perturbative parameters  
because changing the size of $b_{\rm max}$ effectively reshuffles contributions to the TMD PDF between  
the different factors in Eq.~(\ref{eq:evolvedPDF}).
It turns out that 
the $b_{\rm max} = 0.5$~GeV$^{-1}$ value from the BLNY fit is rather small.
In other words, this choice of $b_{\rm max}$ restricts 
the perturbative part of the calculation to a range in $b_T$ that is significantly smaller 
than the range where perturbative methods are still reasonable. 
The analysis in Ref.~\cite{Konychev:2005iy} has 
found that $b_{\rm max} = 1.5$~GeV$^{-1}$ is preferred, and the parameters for the BLNY fit 
in Eq.~(\ref{eq:BLNY}) become $g_1 = 0.201$~GeV$^{2}$, $g_2 = 0.184$~GeV$^{2}$ and $g_3 = -0.129$.
By using the newer parameters from Ref.~\cite{Konychev:2005iy}, 
we can again construct a TMD PDF parametrization from the BLNY form that
matches the STM fit at small $Q$ by using Eq.~(\ref{eq:BLNY3}). 
With the newer parameters we find that $x_0 = 0.009$ is needed to fit to the STM parametrization
at small $k_T$.

One reason that we prefer the smaller $b_{\rm max}$ for the present paper is that
our present analysis includes only the order-$\alpha_s$ contributions to the perturbatively calculable parts, so it is important that  
higher order contributions are small.  In practice, higher order contributions can 
have a large effect, especially at small and intermediate $k_T$.  
We estimate the size of the theoretical error in our analysis 
by redoing the calculation for the parametrization with the larger value of 
$b_{\rm max} = 1.5$~GeV$^{-1}$, and using the parameters of Ref.~\cite{Konychev:2005iy}.
The result is the lower panel in Fig.~\ref{fig:TMDPDF}.  By comparing the upper and lower plots, 
it can be seen that the curves differ by a maximum of about a factor of two 
for the large $Q$ curve.  
(By running the calculation for different 
values of $x$, we have verified that this is generally true for $x$ between about $0.01$ and $0.2$.)
The largest difference is for the $Q = 5$~GeV curves.  This is because
for $Q = 5$~GeV evolution effects become significant, but the different values of $x_0$ needed 
for the curves to match at $Q_0$ lead to a significant difference in $k_T$ dependence.    

In the future, it will be possible to decrease the theoretical uncertainty 
in the TMD PDF parametrization by including higher 
orders in the perturbative parts of the calculation, and by using improved
fits based on newly available data.
  
It is also instructive to investigate the 
effect of the separate A, B and C factors in Eq.~(\ref{eq:evolvedPDF}).
In Fig.~\ref{fig:PDFnoAB}, we have again plotted the TMD PDF for $Q = 91.19$~GeV.
In addition, we show the effect of replacing the $A$ factor by 
simply the lowest order, unevolving result, $f(x,\mu_{b_{\rm max}})$ (blue dashed curve), and 
the effect of replacing the B factor by one (maroon dash-dotted curve).  
(The color version is online.)  
The dashed curves show that simply using $f(x,\mu_{b_{\rm max}})$ instead of the full 
A factor is typically a very good approximation in the small-$k_T$ limit of the TMD PDF.
This significantly simplifies the calculation of the TMD PDF in cases where the very small-$k_T$ region
is the main contribution of interest.
(However, it should be re-emphasized that the full A factor is needed for 
a complete description of the cross section over all $q_T$ up to order $Q$.)
Neglecting the B factor introduces a substantial error at small $k_T$ 
and completely removes the large-$k_T$ tail.
We have compared the calculations for the 
two different $b_{\rm max}$ values, $0.5$~GeV$^{-1}$ and $1.5$~GeV$^{-1}$, in the 
upper and lower plots.  As could be expected, the effect of setting $B = 1$ is substantially larger 
in the $b_{\rm max} = 1.5$~GeV$^{-1}$ case where the role of higher orders is more important.
%%%%%%%%%%%%%%%%%%%%%%%%%%%%%
\begin{figure*}
\centering
\includegraphics[scale=.5]{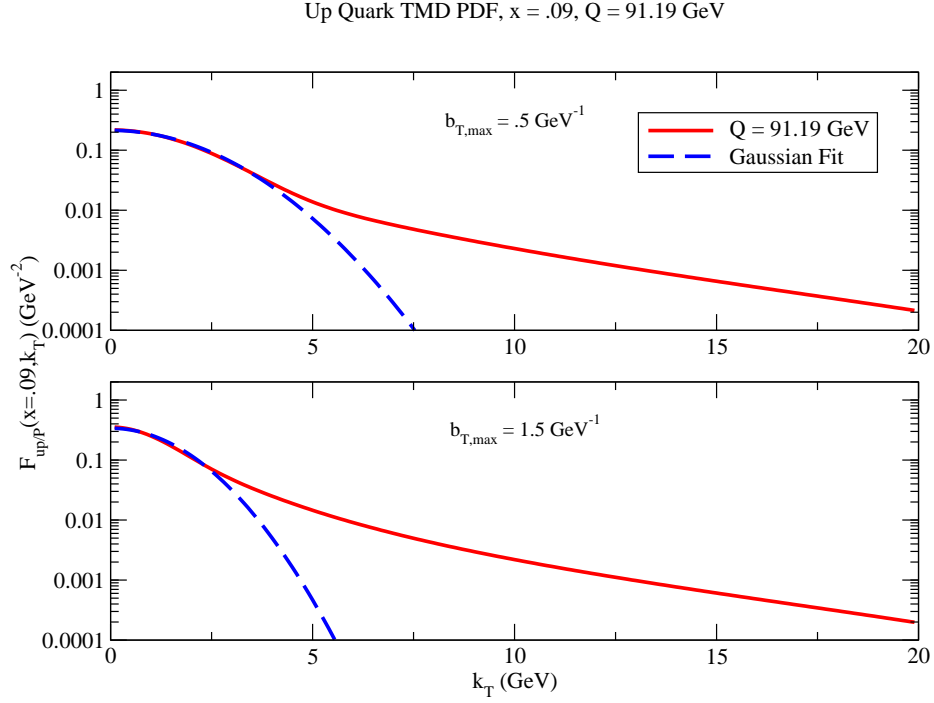}
\caption{Comparison of the TMD PDF at $Q = 91.19$~GeV with a Gaussian fit for the two different values of $b_{\rm max} = 0.5,1.5$~GeV$^{-1}$.
(See online version for color.)
}
\label{fig:gausscomp}
\end{figure*}
%%%%%%%%%%%%%%%%%%%%%%%%%%%%%
%%%%%%%%%%%%%%%%%%%%%%%%%%%%%
\begin{figure*}
\centering
\includegraphics[scale=.5]{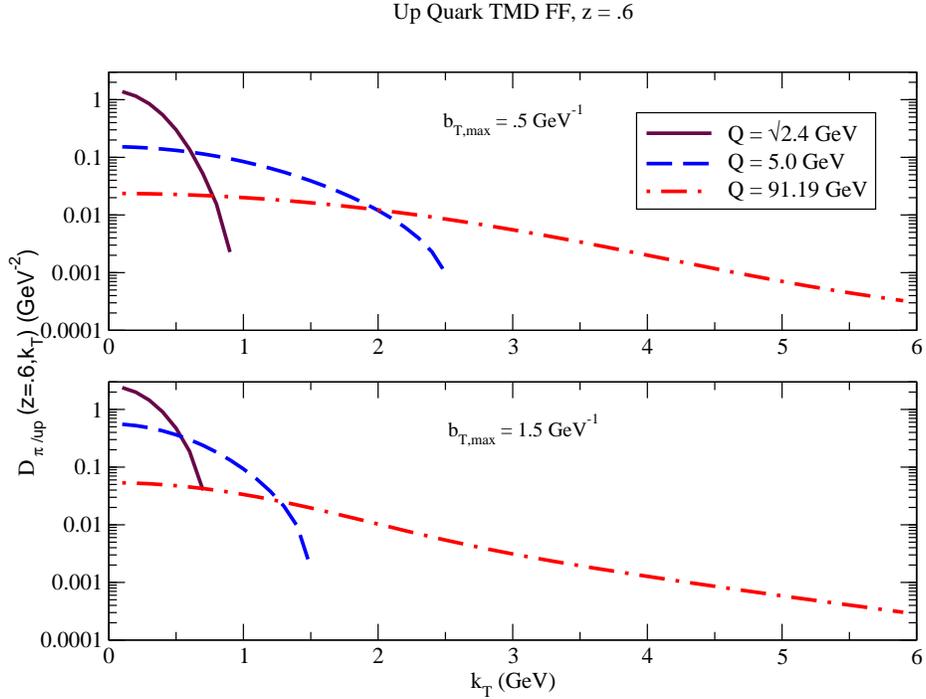}
\caption{TMD FF for charged pions from a hadronizing up-quark.  The upper plot is 
for $g_2$ = $.68$~GeV$^2$ and the lower plot is for $g_2$ = $.184$~GeV$^2$.  In each case, 
TMD FF matches to the STM fit at $Q = \sqrt{2.4}$~GeV.
(See online version for color.)
}
\label{fig:ffscompare}
\end{figure*}
%%%%%%%%%%%%%%%%%%%%%%%%%%%%%

It is common in TMD studies to use Gaussian parametrizations like Eq.~(\ref{eq:STM1}).
An attractive feature of such an approach is that the TMD has a simple and well-defined integral over all $k_T$, and
the standard integrated PDF is obtained simply by integrating the TMD PDF.  
Moreover, the Gaussian form makes the calculation of weighted structure functions simple. 
Therefore, it is useful to investigate how well a Gaussian form describes the shape of the TMD, and 
over what range of transverse momentum. 
(However, we recall that the actual relationship between integrated and TMD PDFs is more complicated, 
as already evidenced by the broad tail in Fig.~\ref{fig:PDFnoAB}.)  
To give an example of such a comparison, and 
to study the effect of the tail 
in fits of the TMD PDF at large $Q$, we have
replotted the $Q = 91.19$~GeV curves from Fig.~\ref{fig:TMDPDF} in Fig.~\ref{fig:gausscomp}, but now we include Gaussian fits.
From the plots it can be seen that at large-$Q$ the Gaussian shape continues to do a reasonable job of describing the 
very small $k_T$ behavior (less than a few GeV), but since it completely neglects the tail at large $k_T$ tail it 
underestimates the size of the typical $k_T$. 

To investigate the role of the tail, 
we calculated 
the integral of the TMD PDF over $k_T$, weighted by $k_T^2$,
\begin{equation}
\label{eq:weighted}
 \bar{k_T^2}  = \int \, d^2 {\bf k}_T \, k_T^2 \, F(x,\trans{k}),
\end{equation} 
and we compared the result of using the Gaussian fit with the result obtained 
by numerically integrating the original $Q = 91.19$~GeV curve (the solid curve in Fig.~\ref{fig:gausscomp}).
For the original curve, Eq.~(\ref{eq:weighted}) is quite ill-defined because 
of the large-$k_T$ tail. 
In addition, the contribution from very large $k_T$ is outside the region where 
a TMD-factorization description alone is valid, 
and the $Y$-term becomes important.  Nevertheless, it is possible to get a sense of the effect of the tail
on typical values of $k_T$ by integrating up to a cutoff that is large, but still significantly less than $Q$.
For the TMD PDFs in Fig.~\ref{fig:gausscomp}, we choose an upper cutoff of $20$~GeV.
For the Gaussian fit, we find $\sqrt{\bar{k_T^2}} = 6$~GeV for $b_{\rm max} = .5$~GeV$^{-1}$ and 
$\sqrt{\bar{k_T^2}} = 4$~GeV for $b_{\rm max} = 1.5$~GeV$^{-1}$.  (The difference 
is due to the slightly different ranges
in $k_T$ where a Gaussian is a good fit.)  For the original 
curve, with the upper cutoff on $k_T$ of $20$~GeV, we find 
$\sqrt{\bar{k_T^2}} = 15$~GeV for both $b_{\rm max} = 0.5$~GeV$^{-1}$ and $b_{\rm max} = 1.5$~GeV$^{-1}$.
Hence, for both values of $b_{\rm max}$, the tail leads to at least 
a factor of two increase in the typical $k_T$.
  
\subsection{TMD FFs}
\label{sec:ffevolv}
The non-perturbative input for the FFs is much less constrained by existing analyses.  
However, the function $g_K(b_T)$ in Eq.~(\ref{eq:evolvedPDF}) for the TMD PDF
is the same function that appears in Eq.~(\ref{eq:evolvedFF}) for the TMD FF. 
Therefore, given a fit for the TMD FF at a particular scale, one can use the same $g_K(b_T)$, 
along with the anomalous dimensions and coefficient functions calculated in Appendices~\ref{sec:coefffuncts},\ref{sec:anomdims}, 
to estimate the evolution to different scales.  
For the starting scale, we again appeal to the fit of Ref.~\cite{Schweitzer:2010tt} which uses a Gaussian form,      
\begin{equation}
\label{eq:STM2}
D_{H/f}(z,{\bf K_T}) = d_{H/f}(z) \times \frac{\exp \left[ -K_T^2 / \langle K_T^2\rangle \right]}{\pi \langle K_T^2\rangle}, 
\end{equation}
where $K_T$ is the hadron transverse momentum
in the photon rest frame.
Again fitting the HERMES data, for 
SIDIS in the kinematical range, $\langle x \rangle = 0.09$, $\langle Q^2 \rangle \approx 2.4$~GeV$^2$ and $z > 0.2$,
they find that 
$\langle K_T^2\rangle = (.16 \pm 0.01)$~GeV$^{2}$. 
One can write the transverse components of the photon-frame hadron momentum ${\bf K}_T$ in terms of 
the transverse components of the hadron-frame parton momentum ${\bf k}_T$ as ${\bf K}_T= -z\,{\bf k}_T$.
The analogue of  Eq.~\eqref{eq:zerothPDF2} for the FF has an extra $1/z^2$ so that 
when all order $\alpha_s$ corrections are dropped the FF reads
\begin{multline}
\label{eq:zerothFF}
\tilde{D}_{H/f}(z,b_T;\zeta_F,\mu) \to \frac{1}{z^2}\,d_{H/f}(z) \\
\times  \exp \left\{ - \left[ g_1^\prime + g_2 z^2 \ln \frac{Q}{ 2 Q_0 } \right] \frac{b_T^2}{2 z^2} \right\}.
\end{multline}
Equating this to the inverse Fourier transform of Eq.~\eqref{eq:STM2}, we identify the factor in brackets as 
$$g_1^\prime + g_2 z^2 \,\ln\frac{\sqrt{2.4 \, {\rm GeV}}}{2 Q_0} \approx \frac{\langle K_T^2\rangle}{2} \approx 0.08 \, {\rm GeV}. $$
From this relation we can extract 
a value for $g_1^\prime$.
The factor multiplying $-b_T^2$ in Eq.~\eqref{eq:zerothFF} can then be identified  
with the non-perturbative exponential factor in Eq.~\eqref{eq:evolvedFF}.
Using Refs.~\cite{frags,Kretzer:2000yf,Kniehl:2000fe,Bourhis:2000gs} for the integrated FFs, 
we can then calculate the TMD FF using Eq.~\eqref{eq:evolvedFF}.
We have repeated the analysis of the TMD PDF for a TMD FF of a charged pion fragmenting from an up quark. 
Fig.~\ref{fig:ffscompare} shows the TMD FF for different energy scales, $Q=\sqrt{2.4}\,\,,5$ and $91.19$ GeV. 
By comparing different energy scales, one can immediately see the effect of including perturbative evolution 
in the definitions of the TMD FFs from the high $k_T$ tails the TMD FFs acquire.  We have also repeated the 
analysis of evaluating the TMD FF for different values of $b_{\rm max}=0.5$ and $1.5$ GeV$^{-1}$ and we find 
a similar error estimate as in the case of the TMD PDF. The comparison is shown again in Fig.~\ref{fig:ffscompare}.
Note that in Fig.~\ref{fig:ffscompare} we have plotted the TMD FF as a function of the \emph{hadron} transverse momentum
$K_T$ rather than parton transverse momentum $k_T$.  

We also investigated how well a Gaussian function fits the perturbatively evolved TMD FF. 
As with the TMD PDF, the Gaussian fit does not adequately capture the effects of perturbative 
evolution for the TMD FF.
The contribution of the $k_T$-tail is smaller in the case of the TMD FF.
This can be understood 
by comparing the $k_T$ dependence of a TMD PDF with a TMD FF.  
The TMD FF is less broad in $k_T$ than a TMD PDF and therefore drops faster with a 
smaller $k_T$ tail. To quantify this we have once more calculated a typical $k_T$ 
using Eq.~\eqref{eq:weighted} both for the Gaussian fit and the actual TMD FF. 
For $b_{\rm max}=0.5$ GeV$^{-1}$ we find that for the Gaussian 
fit $\sqrt{\bar{k_T^2}}=1.74$ GeV while for the actual 
TMD FF $\sqrt{\bar{k_T^2}} = 2.15$ GeV which gives a relative 
difference of  $23.5\%$. For the case of $b_{\rm max}=1.5$ GeV$^{-1}$ the 
values are  $\sqrt{\bar{k_T^2}}=1.06$ GeV for the Gaussian fit 
and  $\sqrt{\bar{k_T^2}} = 1.85$ GeV for the actual TMD FF with a larger relative difference of $73.5\%$.

\section{Discussion and Conclusions}
\label{sec:conclusion}

Factorization theorems provide the bridge between abstract field theoretical 
concepts and phenomenology, and are responsible for giving pQCD its great predictive power.
The parton distribution and fragmentation functions, 
which arise naturally from the factorization derivations, 
play a central role 
in relating formal pQCD to parton model concepts.
A precise understanding of the definitions, evolution, and universality properties of these correlation functions
is what enables calculations in pQCD to make accurate first principles predictions. 

While the standard formalism of collinear factorization has proven  
extremely useful for sufficiently inclusive processes, the more sophisticated 
formalism of TMD-factorization is needed for processes in which  
the intrinsic transverse momentum of the partons becomes important. 
As has already been widely discussed, there are a number of technical and conceptual 
subtleties involved in arriving at good definitions for the TMDs that are consistent with factorization. 
These issues include the need to regulate and deal with rapidity divergences and achieve a cancellation of 
spurious Wilson line self-energies.
The subtleties involved in defining 
TMDs have largely been clarified and resolved in Ref.~\cite{collins}, which provides definitions 
that are consistent with the requirements of factorization, and demonstrates the relationship with the usual 
CSS formalism. 

While  
considerable effort has been 
devoted to implementing CSS evolution in unpolarized scattering, 
the resulting parametrizations are often not framed in the language of TMD PDFs.   
By contrast, for polarization dependent TMDs, there has been very little work 
done in implementing evolution in the parametrization 
of experimental data.  
Up to this point, these functions have only been probed over a very narrow range 
of scales so that evolution has not been a major issue.
However, for future progress in understanding the role of quark and gluon degrees of freedom in  
hadronic structure, it will be  
important
to remedy this situation.  Ideally, there should be collections of tabulated 
fits to the TMDs that incorporate evolution, and which can be directly related to the field-theoretic definitions
of the correlation functions, analogous to what has already existed for some time in collinear factorization.

We have started this process by recasting previously performed 
fits~\cite{Ladinsky:1993zn,Landry:2002ix,Schweitzer:2010tt} of unpolarized TMD PDFs and TMD FFs in terms 
of the TMD definitions of Ref.~\cite{collins}.  
This provides a much clearer connection between the formalism of evolution and 
generalized parton model approaches, and provides practical TMD parametrizations 
that can be used directly in TMD calculations.
We have also completed the derivation 
of the lowest order anomalous dimensions and coefficient functions for the TMD PDF.
At our website~\cite{webpage}, we have supplied tables 
and interpolation routines for the parts of 
the quark TMD PDFs and FFs that can be described 
using collinear factorization (the ``A-factors'' in Eqs.~(\ref{eq:evolvedPDF},~\ref{eq:evolvedFF}))  
for each flavor of quark, as well as 
sample calculations that 
give plots like Fig.~\ref{fig:TMDPDF}.

We have confirmed the important observation that 
evolution has a strong quantitative effect on the TMDs and therefore should be 
included in future phenomenological applications of TMD factorization, particularly 
given the range of energy scales that are set to be probed in the future. Another reason 
to have reliable fits of TMDs, including evolution, is that it opens the possibility to 
identify instances of factorization breaking effects of the type 
discussed in Ref.~\cite{factbreaking1,factbreaking2,factbreaking3,factbreaking4}.
Recognizing factorization breaking effects will be an important next step in expanding
our understanding of pQCD phenomenology.
Even in unpolarized scattering, there is a possibility to use parametrizations like 
those presented in this paper to test the factorization hypothesis.
Recent RHIC data~\cite{Adare:2010bd,Adare:2010yw}, for example, may be useful for 
such an analysis.

Nevertheless, much work remains to be done.  The theoretical uncertainty in the TMD fits can be reduced by 
including higher orders in the calculations of the anomalous dimensions, the $K$ kernel, and the 
collinear coefficient functions.  Moreover, as new data from 
experiments like those taking place at the LHC, RHIC, JLab, and a possible electron-ion collider
are made available and analyzed, it will be possible to obtain improved fits.  
Already, there are data from ATLAS~\cite{:2010yt} 
which can potentially help to 
improve 
the quality of fits for the unpolarized TMDs.
TMD effects can also be studied in an $e^+ e^-$ collider as recently discussed in Ref.~\cite{Boer:2008fr}.
A number of theoretical issues with the evolution formalism itself also remain unsettled. 
For instance, the precise form of the matching function for between perturbative and 
non-perturbative transverse momentum regimes in Eq.~\eqref{bstar} is somewhat arbitrary and 
better prescriptions may be possible.  Along similar lines, a truly optimal choice of $b_{max}$ may be different 
from the values we have used here.  One possiblility may be to formulate the evolution directly in momentum space.

One of the most important
next steps is to extend the analysis 
presented in this paper to the Sivers and Boer-Mulders functions, which are
needed for clarifying the spin structure of hadrons.
Efforts to  
address polarization dependent situations  
can utilize
existing 
fixed-scale fits 
(such as~\cite{Anselmino:2004nk,Anselmino:2005sh,Anselmino:2007fs,Anselmino:2008sga,Anselmino:2009st,Schweitzer:2010tt,D'Alesio:2010am}). 
In such cases, a careful treatment of the matching between large and small transverse momenta will also be important~\cite{Bacchetta:2008xw}.
Furthermore, it will be important to establish the relationship be evolved TMDs and the evolution of weighted functions such as 
those treated in Ref.~\cite{Kang:2010xv}.

Fits of the gluon TMD PDF that include CSS evolution will be also be needed, especially in tests of factorization.  See 
~\cite{Nadolsky:2007ba,Catani:2010pd} for 
recent work related to evolution and 
gluon resummation in the context of gluon PDFs.  
In addition, recent calculations in 
Ref.~\cite{Boer:2010zf} have shown how to probe linearly polarized gluons in heavy quark production,
and the universality properties of the gluon TMD PDF have been clarified in Ref.~\cite{Dominguez:2011wm}.    
For processes that probe the gluon TMDs, 
some important details of the TMD-factorization theorems 
have yet to be completely understood.
The issue of so-called ``super-leading regions'' in the factorization theorems 
that use the gluon distribution~\cite{collinsrogers} still needs to be clarified in the TMD case.
Furthermore, for processes that involve several final state hadrons, such as $e + p \to H_1 + H_2 + X$, the 
separation of the soft factor into universal square root factors as in the definitions
in Eqs.~\eqref{eq:TMDPDFdef} and~\eqref{eq:TMDFFdef} is not straightforward.  Following a naive analysis 
like in Sect.~\ref{sec:intuitive} seems to suggest that 
extra soft factors are needed, and that a more complicated factorization structure is required.

These are all issues we intend to pursue in a continuation of the TMD project.

\begin{acknowledgments} 
We especially thank J.~Collins for many useful discussions regarding his book. 
Helpful comments were also provided by C.~Aidala, D.~Boer, M.~Buffing, A.~Metz, and P.~Mulders. 
We thank P.~Nadolsky for discussions of the BLNY fits, and 
G.~Watt for help in implementing the MSTW parton distribution functions.  
Support was provided by the research 
program of the ``Stichting voor Fundamenteel Onderzoek der Materie (FOM)'', 
which is financially supported by the ``Nederlandse Organisatie voor Wetenschappelijk Onderzoek (NWO)''.
M.~Aybat also acknowledges support from the FP7 EU-programme HadronPhysics2 (contract no 2866403).
All Feynman graphs were made using Jaxodraw~\cite{jaxo}.
\end{acknowledgments}

%%%%%%%%%%%%%%%%%%%%%%%%%%%%%%%%%%%%%%%%%%%%%
\appendix

\section{Coefficient Functions}
\label{sec:coefffuncts}

In this appendix, we present the steps for calculating
the collinear coefficient functions (the 
$\tilde{C}_{f/j}(x/\hat{x},b_{\ast};\mu_b^2,\mu_b,g(\mu_b))$ 
functions in the A-factor of Eq.~\eqref{eq:evolvedPDF}).  We first briefly review the steps, presented in Ref.~\cite{collins}, 
for the case of the TMD FFs.  Then we explain the extension to the 
analogous case for the TMD PDFs.

In perturbation theory, the FF in Eq.~\eqref{eq:FFunsub} itself obeys 
a collinear factorization theorem~\cite{CSS1}, valid for small ${\bf b}_T$. 
The collinear part is just the standard integrated FF.
Writing the factorization as $\tilde{D}=d\otimes C$, we have to first order
\begin{equation}
\tilde{D}^{[1]}=d^{[0]}\otimes \tilde{C}^{[1]} + d^{[1]}\otimes\tilde{C}^{[0]}\,.
\end{equation}
The superscripts label the order in perturbation theory.
Using $d^{[0]}_{j/j^{\prime}}(z) = \delta_{jj^{\prime}}\delta(z-1)$ for the lowest order integrated 
FF, one finds
\begin{equation}
\label{eq:coefFF}
\tilde{C}_{j/f}^{[1]}(z,{\bf b}_T)=\tilde{D}_{j/f}^{[1]}(z,{\bf b}_T)-\frac{d_{j/f}^{[1]}(z)}{z^{2-2\epsilon}}\,,
\end{equation}
for the first order FF coefficient function.  
To get the collinear coefficient function, all that is needed then is to calculate 
the first order expression for the unintegrated FF $\tilde{D}_{j/f}^{[1]}(z,{\bf b}_T)$ and for the 
integrated case $d_{j/f}(z)$.
Since $\tilde{C}_{j/f}^{[1]}(z,{\bf b}_T)$ is independent of the 
species of external hadron, the calculation can be done for the special case of a quark hadronizing to a gluon.

\begin{figure*}
\centering
\includegraphics[scale=0.8]{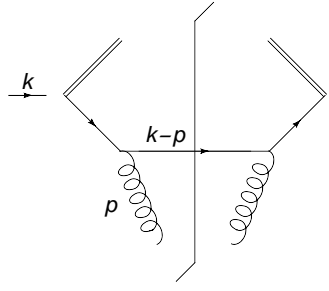}
\caption{One-loop diagram used  
contributing to the TMD FF and the integrated FF of a quark fragmenting to a gluon.}
\label{FFglu}
\end{figure*}

The order $\mathcal{O}(g^2)$ diagram is shown in Fig.~\ref{FFglu}. 
There is no leading contribution from the soft region and hence
no need to subtract soft factor contributions. 
Calculating the first order TMD FF gives
%%%%%%%%%%%%%%%%%%%%%%%%%%%%%
%%%%%%%%%%%%%%%%%%%%%%%%%%%%%
\begin{widetext}
\begin{eqnarray}
\tilde{D}_{g/q}^{[1]}(z,{\bf b}_T) &=& \frac{g^2 \mu^{2\epsilon} C_F}{(2\pi)^{4-2\epsilon}z}\int dk^-d^{2-2\epsilon}{\bf k}_T e^{i{\bf k}_T\cdot{\bf b}_T} 2\pi\delta\left((k-p)^2\right)\frac{1}{4}\frac{{\rm Tr}\sum_j\gamma^+\slash k\gamma^j(\slash k-\slash p )\gamma^j\slash k}{(k^2)^2}\nonumber\\
&=&\frac{g^2(4\pi^2\mu^2)^{\epsilon}C_F}{8\pi^3}\int\frac{d^{2-2\epsilon}{\bf k}_Te^{i{\bf k}_T\cdot{\bf b}_T}}{k_T^2}\Bigg[\frac{1+(1-z)^2-\epsilon z^2}{z^3}\Bigg]\,.
\end{eqnarray}
The corresponding integrated FF is calculated in nearly the same way, except that ${\bf b}_T$ is 
set to zero and the $1/z$ factor in the definition~\eqref{eq:FFunsub} is changed to $z^{2\epsilon-1}$ in 
the integrated case.  Also, an $\overline{\rm MS}$ counterterm is needed to remove the 
resulting UV divergence.
The result is
\begin{equation}
d_{g/q}^{[1]}(z)=\frac{g^2(4\pi^2\mu^2)^{\epsilon}C_F}{8\pi^3}\int\frac{d^{2-2\epsilon}{\bf k}_T}{k_T^2}\Bigg[\frac{1+(1-z)^2-\epsilon z^2}{z^{1+2\epsilon}}\Bigg] - \frac{g^2 C_F (4\pi)^{\epsilon}}{8\pi^2\Gamma(1-\epsilon)\epsilon}\Bigg[\frac{1+(1-z)^2}{z}\Bigg]\,.
\end{equation}
Performing the $k_T$ integrals and putting all the terms together in Eq.~\eqref{eq:coefFF} gives the collinear coefficient function
\begin{equation}
\label{eq:collcoefgq}
\tilde{C}_{g / j^\prime}(z,\trans{b};\mu;\zeta_D / \mu^2) = \frac{\alpha_s C_{\rm F}}{2 \pi z^3} \left( 2 \left[ 1 + (1 - z)^2 \right] 
\left[ \ln \left( \frac{2 z}{\subT{b} \mu} \right) - \gammae \right] + z^2 \right) + \mathcal{O}(\alpha_s^2)\,.
\end{equation}
%%%%%%%%%%%%%%%%%%%%%%%%%%%%%
\begin{figure*}
\centering
\includegraphics[scale=0.8]{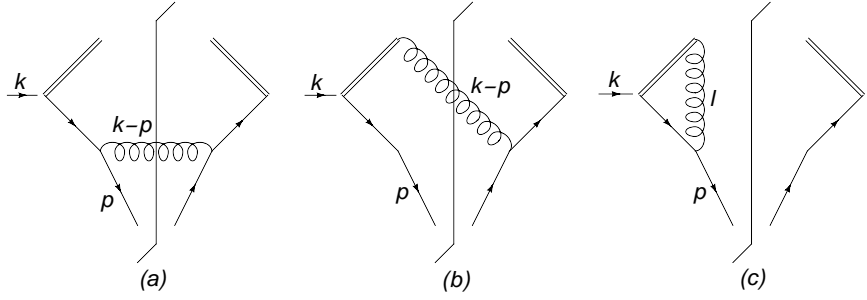}
\caption{One-loop diagrams 
contributing to
the TMD FF and the integrated FF of a quark fragmenting into a quark.}
\label{FFquark}
\end{figure*}
%%%%%%%%%%%%%%%%%%%%%%%%%%%%%
%%%%%%%%%%%%%%%%%%%%%%%%%%%%%
\begin{figure*}
\centering
\includegraphics[scale=0.6]{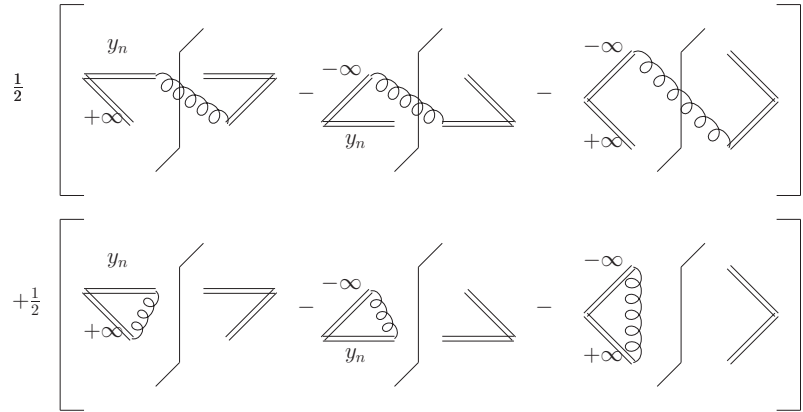}
\caption{One-loop diagrams for the soft-factor contributions of Eq.~\eqref{eq:evolvedFF}.
Hermitian conjugate graphs are also needed but are not shown.
}
\label{soft}
\end{figure*}
%%%%%%%%%%%%%%%%%%%%%%%%%%%%%

For the TMD FF of a quark hadronizing to a quark, the diagrams are shown in Fig.~\ref{FFquark}, along 
with the soft subtraction factors in Fig.~\ref{soft} that are needed according to the definition in Eq.~\eqref{eq:TMDPDFdef}. 
Apart from the need to include soft factor contributions, the steps are analogous to those that 
led to Eq.~\eqref{eq:collcoefgq}
The integrated FF is found 
again using Fig.~\ref{FFquark} along with $\overline{\rm MS}$ counterterms. The result is
\begin{multline}
\tilde{C}_{j / j^\prime}(z,\trans{b};\mu;\zeta_D / \mu^2) = \delta_{j^\prime j} \delta(1 - z) +  \delta_{j^\prime j} \frac{\alpha_s C_{\rm F}}{2 \pi} 
\left\{ 2 \left[ \ln \left( \frac{2 z}{\mu \subT{b}} \right) - \gammae  \right] 
\left[ \left( \frac{2}{1 - z} \right)_{+} + \frac{1}{z^2} + \frac{1}{z} \right] + \frac{1}{z^2} - \frac{1}{z} \right. + \\ 
\left. + \delta(1 - z) \left[  - \frac{1}{2} \left[ \ln \left( \subT{b}^2 \mu^2 \right) - 2(\ln 2 - \gammae) \right]^2  - \left[\ln(\subT{b}^2 \mu^2) 
- 2 (\ln 2 - \gammae ) \right] \ln \left( \frac{\zeta_D}{\mu^2} \right) \right] \right\} + \mathcal{O}(\alpha_s^2).
\end{multline}
%%%%%%%%%%%%%%%%%%%%%%%%%%%%%
\begin{figure*}
\centering
\includegraphics[scale=0.8]{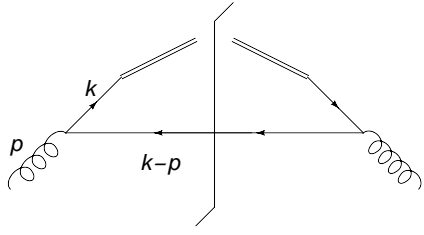}
\caption{One-loop diagram  
contributing to the TMD 
PDF
and the integrated PDF for a quark inside a gluon.}
\label{PDFglu}
\end{figure*}
%%%%%%%%%%%%%%%%%%%%%%%%%%%%%
%%%%%%%%%%%%%%%%%%%%%%%%%%%%%
\begin{figure*}
\centering
\includegraphics[scale=1.0]{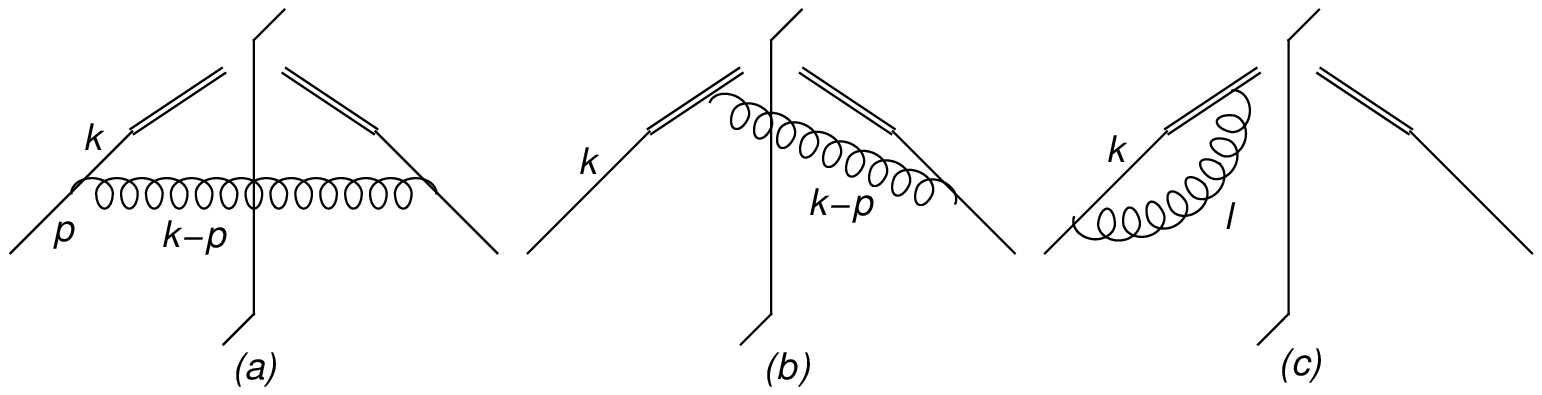}
\caption{One-loop diagrams 
contributing to the TMD PDF and the integrated PDF of a quark inside a quark.
Hermitian conjugate graphs are also needed but are note shown.}
\label{PDFquark}
\end{figure*}
%%%%%%%%%%%%%%%%%%%%%%%%%%%%%
Again, the details for the above FF calculation can be found in Ref.~\cite{collins}.

The steps for calculating the 
order-$\mathcal{O}(g^2)$ contributions to the collinear coefficient functions 
for the TMD PDFs are analogous to the steps used in the TMD FF case, with some 
minor changes.  
We provide them here, presented for the first time in the context 
of TMDs.  The results should be equivalent to calculations already done in the CSS formalism, up 
to possible changes in scheme.
For the TMD PDF, the analogue of Eq.~\eqref{eq:coefFF} is
\begin{equation}
\label{PDFconv}
\tilde{C}_{j/f}^{[1]}(x,{\bf b}_T)=\tilde{F}_{j/f}^{[1]}(x,{\bf b}_T)-f_{j/f}^{[1]}(x)\,.
\end{equation}
The difference in factors of longitudinal momentum fraction in the second term comes from the 
different normalization in Eq.~\eqref{eq:smallbFF} as compared with~\eqref{eq:smallb}.
Again, we may perform the calculation for on-shell external partons.
Using the diagram in Fig.~\ref{PDFglu}, for the TMD PDF of a quark inside a gluon one has
\begin{eqnarray}
\label{TMDunintegratedPDF}
\tilde{F}_{q/g}^{[1]}(x,{\bf b}_T) & = & -\frac{T_f g^2 \mu^{2\epsilon}}{(2\pi)^{4-2\epsilon}(1-\epsilon)}
\int\,dk^-d^{2-2\epsilon}{\bf k}_T\,e^{-i{\bf k}_T\cdot{\bf b}_T}2\pi\delta\left((p-k)^2\right)
\frac{1}{4}\frac{\sum_{\rm pol}{\rm Tr}(\gamma^+\slash k \slash \epsilon_p (\slash k - \slash p) \slash \epsilon_p^* \slash k)}{(k^2)^2} \nonumber \\
& = & \frac{g^2 (4\pi\mu^2)^{\epsilon}T_f}{8\pi^2\Gamma(1-\epsilon)}
\int_0^{\infty} \frac{d^{2-2\epsilon}{\bf k}_T}{k_T^2} e^{-i {\bf k}_T \cdot {\bf b}_T } \Bigg[1-\frac{2x(1-x)}{1-\epsilon}\Bigg
]
\end{eqnarray}
where $\epsilon_p^{\mu}$ is the polarization vector for the initial state 
gluon and we sum and average over all possible polarizations. The integrated 
PDF is found again by setting ${\bf b}_T=0$ in the above equation and adding 
the appropriate $\overline{\rm MS}$ counterterm for the resulting UV divergence. This gives
\begin{equation}
\label{TMDintegratedPDF}
f_{q/g}^{[1]}(x)= \frac{g^2 (4\pi\mu^2)^{\epsilon}T_f}{8\pi^2\Gamma(1-\epsilon)}
\int_0^{\infty}dk_T^2\frac{k_T^{-2\epsilon}}{k_T^2}\Bigg[1-\frac{2x(1-x)}{1-\epsilon}\Bigg
]-\frac{g^2(4\pi\mu^2)^{\epsilon}T_f}{8\pi^2\Gamma(1-\epsilon)\epsilon}\Bigg[1-2x(1-x)\Bigg]\,. 
\end{equation}
Using Eqs.\eqref{PDFconv}-\eqref{TMDintegratedPDF} and evaluating the $k_T$ integrals, 
gives the TMD PDF collinear coefficient function 
for finding a quark of flavor $j^\prime$ in a gluon at order $\alpha_s$,
\begin{equation}
\tilde{C}_{j^\prime / g}(x,\trans{b};\mu;\zeta_F / \mu^2) =    \frac{\alpha_s T_{\rm f}}{2 \pi} \left( 2 \left[ 1 - 2x(1 - x) \right] 
\left[ \ln \left( \frac{2}{\subT{b} \mu} \right) - \gammae \right] + 2 x(1 - x) \right) + \mathcal{O}(\alpha_s^2)
\end{equation}

Finally using diagrams in Fig.~\ref{PDFquark} together with the soft 
subtraction terms in Fig.~\ref{soft} for the TMD PDF for finding a 
quark  of flavor $j^\prime$ in a quark of flavor $j$ and again the 
diagrams in Fig.~\ref{PDFquark} together with the $\overline{\rm MS}$ 
UV counterterms for the integrated quark PDF we find to order $\alpha_s$,
\begin{multline}
\tilde{C}_{j^\prime / j}(x,\trans{b};\mu;\zeta_F / \mu^2) = \delta_{j^\prime j} \delta(1 - x) +  \delta_{j^\prime j} \frac{\alpha_s C_{\rm F}}{2 \pi} 
\left\{ 2 \left[ \ln \left( \frac{2}{\mu \subT{b}} \right) - \gammae  \right] 
\left[ \left( \frac{2}{1 - x} \right)_{+} - 1 - x \right] + 1 - x \right. + \\ 
\left. + \delta(1 - x) \left[  - \frac{1}{2} \left[ \ln \left( \subT{b}^2 \mu^2 \right) - 2(\ln 2 - \gammae) \right]^2  - \left[\ln(\subT{b}^2 \mu^2) 
- 2 (\ln 2 - \gammae ) \right] \ln \left( \frac{\zeta_F}{\mu^2} \right) \right] \right\} + \mathcal{O}(\alpha_s^2).
\end{multline}

\end{widetext}

\section{Anomalous Dimensions}
\label{sec:anomdims}
All calculations of anomalous dimensions defined 
in Eqs.~\eqref{eq:RGKPDF},\eqref{eq:RGPDF} and \eqref{eq:RGFF} 
use dimensional regularization with the $\overline{{\rm MS}}$ scheme.
The anomalous dimension of the quark TMD PDF up to order $\alpha_s$ is,
\begin{equation}
\label{eq:anom}
\gamma_{\rm F}(\mu;\zeta_F / \mu^2) = \alpha_s  \frac{C_{\rm F}}{\pi} \left(\frac{3}{2} - \ln \left( \frac{\zeta_F}{\mu^2} \right) \right)
+ \mathcal{O}(\alpha_s^2).
\end{equation}
At order $\alpha_s$, the quark TMD FF anomalous dimension is the same as for the TMD PDF.
We note that these results are consistent with what is found in, e.g., Ref.~\cite{Stefanis:2010vu} using different methods.

The CS kernel, in Eq.~(\ref{eq:CSPDF}), up to order $\alpha_s$ in ${\bf b}_T$-space is,
\begin{equation}
\label{eq:kern}
\tilde{K}(\mu,b_T) = - \frac{\alpha_s C_F}{\pi} \left[ \ln(\mu^2 b_T^2) - \ln 4 + 2 \gammae \right]
+ \mathcal{O}(\alpha_s^2).
\end{equation}
The anomalous dimension of $\tilde{K}$ (see Eq.~(\ref{eq:RGKPDF})) is up to order $\alpha_s$,
\begin{equation}
\label{eq:Kanom}
\gamma_K(\mu) = 2 \frac{\alpha_s C_F}{\pi} + \mathcal{O}(\alpha_s^2).
\end{equation}

\end{document}